\documentstyle[12pt,epsfig]{article}

\newskip\humongous \humongous=0pt plus 1000pt minus 1000pt

\newif\ifdtup

\def\abs#1{\left| #1\right|}
\def\pr#1{#1^\prime}

\def\beq{\begin{equation}}
\def\eeq{\end{equation}}

\def\beqn{\begin{eqnarray}}
\def\eeqn{\end{eqnarray}}
\relax

\def\dotx{\dotx{\dot\overline{x}}}

\relax

\catcode`\@=11

\@addtoreset{equation}{section}
\def\theequation{\thesection\arabic{equation}}

\def\@normalsize{\@setsize\normalsize{15pt}\xiipt\@xiipt
\abovedisplayskip 14pt plus3pt minus3pt%
\belowdisplayskip \abovedisplayskip
\abovedisplayshortskip \z@ plus3pt%
\belowdisplayshortskip 7pt plus3.5pt minus0pt}

\def\small{\@setsize\small{13.6pt}\xipt\@xipt
\abovedisplayskip 13pt plus3pt minus3pt%
\belowdisplayskip \abovedisplayskip
\abovedisplayshortskip \z@ plus3pt%
\belowdisplayshortskip 7pt plus3.5pt minus0pt
\def\@listi{\parsep 4.5pt plus 2pt minus 1pt
     \itemsep \parsep
     \topsep 9pt plus 3pt minus 3pt}}

\@twosidetrue

\relax

\catcode`@=12

\catcode`\@=11

\def\section{\@startsection{section}{1}{\z@}{3.5ex plus 1ex minus
   .2ex}{2.3ex plus .2ex}{\large\bf}}

\def\thesection{\arabic{section}.}

\def\appendix{\setcounter{section}{0}
 \def\thesection{APPENDIX \Alph{section}:}
 \def\theequation{\Alph{section}.\arabic{equation}}}

\def\ps@headings{\def\@oddfoot{}\def\@evenfoot{}
\def\@oddhead{\hbox{}\hfill
 \makebox[.5\textwidth]{\raggedright\ignorespaces --\thepage{}--
 \hfill {}}}  
\def\@evenhead{\@oddhead}
\def\subsectionmark##1{\markboth{##1}{}}
}

\ps@headings

\catcode`\@=12

\def\figcap{\section*{Figure Captions\markboth
 {FIGURECAPTIONS}{FIGURECAPTIONS}}\list
 {Fig. \arabic{enumi}:\hfill}{\settowidth\labelwidth{Fig. 999:}
 \leftmargin\labelwidth
 \advance\leftmargin\labelsep\usecounter{enumi}}}
 \relax
\def\tablecap{\section*{Table Captions\markboth
 {TABLECAPTIONS}{TABLECAPTIONS}}\list
 {Table \arabic{enumi}:\hfill}{\settowidth\labelwidth{Table 999:}
 \leftmargin\labelwidth
 \advance\leftmargin\labelsep\usecounter{enumi}}}
 \relax
\def\reflist{\section*{References\markboth
 {REFLIST}{REFLIST}}\list
 {[\arabic{enumi}]\hfill}{\settowidth\labelwidth{[999]}
 \leftmargin\labelwidth
 \advance\leftmargin\labelsep\usecounter{enumi}}}
 \relax

\catcode`\@=11

\def\ps@headings{\def\@oddfoot{}\def\@evenfoot{}
\def\@oddhead{\hbox{}\hfill
 \makebox[.5\textwidth]{\raggedright\ignorespaces --\thepage{}--
 \hfill {}}}    
\def\@evenhead{\@oddhead}
\def\subsectionmark##1{\markboth{##1}{}}
}

\ps@headings

\catcode`\@=12

\relax

\catcode`\@=11
\def\prm{\fam \z@}
\catcode`\@=12

\relax
\def\pl#1#2#3{{\it Phys. Lett. }{\bf #1}(19#2)#3}
\def\zp#1#2#3{{\it Z. Phys. }{\bf #1}(19#2)#3}
\def\prl#1#2#3{{\it Phys. Rev. Lett. }{\bf #1}(19#2)#3}

\def\pr#1#2#3{{\it Phys. Rev. }{\bf #1}(19#2)#3}
\def\np#1#2#3{{\it Nucl. Phys. }{\bf #1}(19#2)#3}

\def    \hepph  #1 {{\tt hep-ph/#1}}
\def    \hepex  #1 {{\tt hep-ex/#1}}

\relax

  \newcommand{\ccaption}[2]{
    \begin{center}
    \parbox{0.85\textwidth}{
      \caption[#1]{\small{\it{#2}}}
      }
    \end{center}
    }
\begin{document}            
\newcommand\sss{\scriptscriptstyle}
\newcommand\mug{\mu_\gamma}
\newcommand\mue{\mu_e}
\newcommand\muf{\mu_{\sss F}}
\newcommand\mufp{\mu_{\sss F}^\prime}
\newcommand\mufs{\mu_{\sss F}^{\prime\prime}}
\newcommand\mur{\mu_{\sss R}}
\newcommand\murp{\mu_{\sss R}^\prime}
\newcommand\murs{\mu_{\sss R}^{\prime\prime}}
\newcommand\muh{\mu_{\sss H}}
\newcommand\muhp{\mu_{\sss H}^\prime}
\newcommand\muhs{\mu_{\sss H}^{\prime\prime}}
\newcommand\muo{\mu_0}
\newcommand\etonecut{E_{1\sss T}^{cut}}
\newcommand\ettwocut{E_{2\sss T}^{cut}}
\newcommand\etone{E_{1\sss T}}
\newcommand\ettwo{E_{2\sss T}}
\newcommand\me{m_e}
\newcommand\as{\alpha_{\sss S}}         
\newcommand\ep{\epsilon}
\newcommand\Th{\theta}
\newcommand\epb{\overline{\epsilon}}
\newcommand\aem{\alpha_{\rm em}}
\newcommand\refq[1]{$^{[#1]}$}
\newcommand\avr[1]{\left\langle #1 \right\rangle}
\newcommand\lambdamsb{\Lambda_5^{\rm \sss \overline{MS}}}
\newcommand\qqb{{q\overline{q}}}
\newcommand\qb{\overline{q}}
\newcommand\et{E_{\sss T}}
\newcommand\etcut{E^{cut}_{\sss T}}
\newcommand\MSB{{\rm \overline{MS}}}
\newcommand\DIG{{\rm DIS}_\gamma}
\newcommand\CA{C_{\sss A}}
\newcommand\DA{D_{\sss A}}
\newcommand\CF{C_{\sss F}}
\newcommand\TF{T_{\sss F}}
\newcommand\Jetlist{\{J_l\}_{1,3}}
\newcommand\aoat{\sss a_{\sss 1} a_{\sss 2}}
\newcommand\SFfull{\{k_l\}_{1,6}}
\newcommand\SFfullbi{\{k_l\}_{1,6}^{[i]}}
\newcommand\SFTfull{\{k_l\}_{1,5}}
\newcommand\SFj{\{k_l\}_{3,6}}
\newcommand\FLfull{\{a_l\}_{1,6}}
\newcommand\FLfullbi{\{a_l\}_{1,6}^{[i]}}
\newcommand\FLTfull{\{a_l\}_{1,5}}
\newcommand\FLFj{\{a_l\}_{3,6}}
\newcommand\SFjbi{\{k_l\}_{3,6}^{[i]}}
\newcommand\FLjbi{\{a_l\}_{3,6}^{[i]}}
\newcommand\SFjexcl{\{k_l\}_{i,j}^{[np..]}}
\newcommand\SCollfull{\{k_l\}_{1,7}^{[ij]}}
\newcommand\SColl{\{k_l\}_{3,7}^{[ij]}}
\newcommand\FLCollfull{\{a_l\}_{1,7}^{[ij]}}
\newcommand\FLColl{\{a_l\}_{3,7}^{[ij]}}
\newcommand\STj{\{k_l\}_{3,5}}
\newcommand\FLTj{\{a_l\}_{3,5}}
\newcommand\FLNj{\{a_l\}_{3}^{N+2}}
\newcommand\FLNmoj{\{a_l\}_{3}^{N+1}}
\newcommand\FLNfullj{\{a_l\}_{1}^{N+2}}
\newcommand\FLNmofullj{\{a_l\}_{1}^{N+1}}
\newcommand\Argfull{\FLfull;\SFfull}
\newcommand\ArgTfull{\FLTfull;\SFTfull}
\newcommand\KtoKF{(k_1,k_2\to\SFj\,;\FLFj)}
\newcommand\KtoKT{(k_1,k_2\to\STj\,;\FLTj)}
\newcommand\FLsum{\sum_{\{a_l\}}}
\newcommand\FLFsum{\sum_{\FLFj}}
\newcommand\FLFsumbi{\sum_{\FLjbi}}
\newcommand\FLTsum{\sum_{\FLTj}}
\newcommand\FLNsum{\sum_{\FLNj}}
\newcommand\FLNmosum{\sum_{\FLNmoj}}
\newcommand\MF{{\cal M}^{(4)}}
\newcommand\MT{{\cal M}^{(3)}}
\newcommand\MTz{{\cal M}^{(3,0)}}
\newcommand\MTo{{\cal M}^{(3,1)}}
\newcommand\MTi{{\cal M}^{(3,i)}}
\newcommand\MTmn{{\cal M}^{(3,0)}_{mn}}
\newcommand\MN{{\cal M}^{(N)}}
\newcommand\MNmo{{\cal M}^{(N-1)}}
\newcommand\MNmoij{{\cal M}^{(N-1)}_{ij}}
\newcommand\MNmoV{{\cal M}^{(N-1,v)}}
\newcommand\MFsj{\MF(k_1,k_2\to\SFj)}
\newcommand\MTsj{\MT(k_1,k_2\to\STj)}
\newcommand\MTisj{\MTi(k_1,k_2\to\STj)}
\newcommand\PHIFsj{\phi_4(k_1,k_2\to\SFj)}
\newcommand\PHITsj{\phi_3(k_1,k_2\to\STj)}
\newcommand\uoffct{\frac{1}{4!}}
\newcommand\uotfct{\frac{1}{3!}}
\newcommand\uoNfct{\frac{1}{N!}}
\newcommand\uoNmofct{\frac{1}{(N-1)!}}
\newcommand\uoxic{\left(\frac{1}{\xi}\right)_c}
\newcommand\uoxiic{\left(\frac{1}{\xi_i}\right)_c}
\newcommand\uoxilc{\left(\frac{\log\xi}{\xi}\right)_c}
\newcommand\uoxiilc{\left(\frac{\log\xi_i}{\xi_i}\right)_c}
\newcommand\uoyim{\left(\frac{1}{1-y_i}\right)_+}
\newcommand\uoyimdi{\left(\frac{1}{1-y_i}\right)_{\delta_{\sss I}}}
\newcommand\uoyimpdi{\left(\frac{1}{1\mp y_i}\right)_{\delta_{\sss I}}}
\newcommand\uoyjmdo{\left(\frac{1}{1-y_j}\right)_{\delta_o}}
\newcommand\uoyip{\left(\frac{1}{1+y_i}\right)_+}
\newcommand\uoyipdi{\left(\frac{1}{1+y_i}\right)_{\delta_{\sss I}}}
\newcommand\uoyilm{\left(\frac{\log(1-y_i)}{1-y_i}\right)_+}
\newcommand\uoyilp{\left(\frac{\log(1+y_i)}{1+y_i}\right)_+}
\newcommand\uozm{\left(\frac{1}{1-z}\right)_+}
\newcommand\uozlm{\left(\frac{\log(1-z)}{1-z}\right)_+}
\newcommand\SVfact{\frac{(4\pi)^\ep}{\Gamma(1-\ep)}
                   \left(\frac{\mu^2}{Q^2}\right)^\ep}
\newcommand\gs{g_{\sss S}}
\newcommand\Icol{\{d_l\}}
\newcommand\An{{\cal A}^{(n)}}
\newcommand\Mn{{\cal M}^{(n)}}
\newcommand\Nn{{\cal N}^{(n)}}
\newcommand\Anu{{\cal A}^{(n-1)}}
\newcommand\Mnu{{\cal M}^{(n-1)}}
\newcommand\Sumae{\sum_{d_{e}}}
\newcommand\Sumaecl{\sum_{d_{e},\Icol}}
\newcommand\Sumaeae{\sum_{d_{e},d_{e}^{\prime}}}
\newcommand\Sumhe{\sum_{h_{e}}}
\newcommand\Sumhep{\sum_{h_{e}^\prime}}
\newcommand\Sumhehe{\sum_{h_{e},h_{e}^{\prime}}}
\newcommand\Pgghhh{S_{gg}^{h_e h_i h_j}}
\newcommand\Pggplus{S_{gg}^{+ h_i h_j}}
\newcommand\Pggminus{S_{gg}^{- h_i h_j}}
\newcommand\Pgghphh{S_{gg}^{h_e^\prime h_i h_j}}
\newcommand\Pqgplus{S_{qg}^{+ h_i h_j}}
\newcommand\Pqgminus{S_{qg}^{- h_i h_j}}
\newcommand\Pgqplus{S_{gq}^{+ h_i h_j}}
\newcommand\Pgqminus{S_{gq}^{- h_i h_j}}
\newcommand\Pqqplus{S_{qq}^{+ h_i h_j}}
\newcommand\Pqqminus{S_{qq}^{- h_i h_j}}
\newcommand\Hnd{\{h_l\}}
\newcommand\LC{\stackrel{\sss i\parallel j}{\longrightarrow}}
\newcommand\LCu{\stackrel{\sss 1\parallel j}{\longrightarrow}}
\newcommand\Physvar{\{V_l\}}
\newcommand\Physcm{\{\bar{V}_l\}}
\newcommand\Partvar{\{v_l\}}
\newcommand\Partcm{\{\bar{v}_l\}}
\newcommand\Partvarcm{\{v_l(\bar{v})\}}
\newcommand\Partcmset{\{\bar{v}_l^{(i)}\}}
\newcommand\Partvarcmset{\{v_l(\bar{v}^{(i)})\}}
\def\A#1#2{\la#1#2\ra} 
\def\B#1#2{[#1#2]} 
\newcommand{\la}{\langle}
\newcommand{\ra}{\rangle}
\newcommand{\nn}{\nonumber}
\newcommand\Qb{\overline{Q}}
\renewcommand\topfraction{1}       
\renewcommand\bottomfraction{1}    
\renewcommand\textfraction{0}      
\setcounter{topnumber}{5}          
\setcounter{bottomnumber}{5}       
\setcounter{totalnumber}{5}        
\setcounter{dbltopnumber}{2}       
\newsavebox\tmpfig
\newcommand\settmpfig[1]{\sbox{\tmpfig}{\mbox{\ref{#1}}}}
%
\begin{titlepage}
\nopagebreak
\flushright{
        \begin{minipage}{4cm}
        ETH-TH/97-21 \hfill \\
        GEF-TH-8/1997 \hfill \\
        hep-ph/9707345\hfill \\
        \end{minipage}        }

\vfill
\begin{center}
{\large\sc Jet photoproduction at HERA}
\vskip .5cm
{\bf Stefano FRIXIONE}\footnote{Work supported by the Swiss National
Foundation.}
\\                    
\vskip .1cm
{Theoretical Physics, ETH, Zurich, Switzerland} \\
\vskip .5cm                                      
{\bf Giovanni RIDOLFI}
\\
\vskip 0.1cm
{INFN, Sezione di Genova, Genoa, Italy} \\
\end{center}
\nopagebreak
\vfill
\begin{abstract}
We compute various kinematical distributions for one-jet and two-jet
inclusive photoproduction at HERA. Our results are accurate to
next-to-leading order in QCD. We use the subtraction method
for the cancellation of infrared singularities. We perform a 
thorough study of the reliability of QCD predictions; in particular,
we consider the scale dependence of our results and discuss the
cases when the perturbative expansion might break down.
We also deal with the problem of the experimental definition of 
the pointlike and hadronic components of the incident photon,
and briefly discuss the sensitivity of QCD predictions upon the input
parameters of the calculation, like $\as$ and the parton densities.

\end{abstract}        
\vfill
\end{titlepage}
\section{Introduction}
Jet production is a frequent phenomenon in high-energy collisions.
It is characterized by large rates, which allow measurements of its 
kinematical features even in the case of high jet multiplicity (like 
six-jet events at Tevatron and five-jet events at LEP).
Complete next-to-leading order QCD calculations are available
for one- and two-jet inclusive quantities in hadronic 
collisions~[\ref{NLOhadcoll}], and up to four-jet inclusive quantities
in $e^+e^-$ collisions~[\ref{NLOepemcoll}]. The typical energy scale for the
hard process is of the order of the average jet
transverse momentum; at presently operating colliders, this means that
the hard production process takes place in an energy domain
where the predictions of perturbative QCD are particularly accurate.

At hadron colliders, for one- and two-jet inclusive observables the 
uncertainty affecting the theoretical results is smaller than experimental
errors, and stringent tests for QCD can be carried out. Although
most of the comparisons between theory and data are quite
successful, some issues need to be better clarified. In particular,
we recall that the CDF collaboration
reports~[\ref{CDFhighet}] an excess of events
in the tail of the $\et$ distribution with respect to the
theoretical predictions (this discrepancy, however, can be absorbed in a 
suitable modification of the parton densities~[\ref{CTEQ4}]),
and that the comparison between data~[\ref{scaling}] taken at 
different center-of-mass energies in $p\bar{p}$ collisions
is still not consistent with the scaling found in QCD.

The amount of experimental information relevant for lepton-hadron
collisions is presently not as detailed as in the case of
hadronic or $e^+e^-$ processes. Nevertheless, in the near future
the increased luminosity of the HERA collider will allow a
statistically significant study also for this kind of hard scattering.
In jet production at HERA, one can consider two different kinematical
regimes. In one class of events, the electron is scattered at large
angles (deep inelastic scattering). In another
class of events, the electron is lost in the beam-pipe (or tagged
at very small angles); in this case, it acts as a source of real
photons, which eventually interact with the partons in the proton
beam. It is well known that an on-shell photon has a finite probability
of fluctuating into a hadronic state before undergoing a hard collision.
Therefore, for a comparison with experimental data on jet photoproduction, 
one must consider QCD processes in which the photon directly
enters the hard parton scattering, and processes in which the photon
emits partons (in a way parametrized by some universal but non-calculable
parton densities) which enter the hard parton scattering.

Partial next-to-leading order QCD results for jet photoproduction
have been available for some time~[\ref{NLOatHERAnc}]. Complete
calculations for one-jet and two-jet inclusive quantities which include
the contributions of all the relevant partonic subprocesses have
been reported in refs.~[\ref{KKfull}-\ref{mcjet}]. Comparison with data 
are encouraging (see for example refs.~[\ref{Butterworth},\ref{Aurenche}]
for a recent review on this point), but larger statistics is needed
in order to study more exclusive quantities and to fully
test the predictions of the theory. Therefore, the aim of this paper
is not to perform a comparison with existing data,
but rather to discuss some general features of the theoretical predictions
of jet photoproduction in the HERA energy range.
Our predictions are obtained using the computer codes~\footnote{The codes,
which return parton kinematical configurations with an appropriate weight,
are available upon request.} recently presented in ref.~[\ref{mcjet}],
and are based on the general formalism of ref.~[\ref{FKS}].
The fundamental difference between the results of 
refs.~[\ref{KKfull},\ref{HO}]
and those presented here is that the former are obtained
in the framework of the slicing method, while the latter rely upon
the subtraction method. Both methods have been devised in order to
cancel analytically the infrared divergencies which arise in the
intermediate steps of any next-to-leading order QCD calculation.
The slicing method requires the matrix elements for the 
real emission processes to be approximated in those regions of the
phase space which are close to an infrared (soft or collinear)
singularity. These
regions are defined by means of some non-physical parameters.
Physical results must be
independent of these parameters; it is in general a non-trivial task 
to prove (numerically) that this is indeed the case, at the required
level of accuracy. Notice also that in 
principle this proof has to be carried out for each observable, since 
different quantities may display different convergence properties.
In the subtraction method, the exact expression of the matrix elements
is used in the whole phase space. The main advantage of this 
method is that it does {\it not} require the introduction of any 
non-physical parameter. In the following, we will encounter cases where this 
feature of the subtraction method will turn out to be advantageous
in the calculation of physical quantities.

This paper is organized as follows. In section 2 we study the problem
of the reliability of the next-to-leading order QCD predictions
for jet observables at HERA, where the typical transverse momenta
are sizeably smaller than in the case of hadron collider experiments.
We consider several one-jet and two-jet inclusive quantities,
and we isolate the regions where QCD perturbative expansion breaks
down, and an all-order resummation of soft-gluon effects is
mandatory. We also discuss the problem of the {\it operational}
separation of the pointlike (or direct) and of the hadronic
(or resolved) components of the cross section. In section 3
we present some predictions for quantities whose measurement
would help the understanding of the jet production mechanism and 
ultimately to better test QCD in the context of photon-hadron
collisions. Finally, in section 4 we report our conclusions.

\section{General features of QCD predictions}
Any production cross section in electron-proton collisions is dominated
by the exchange of low-virtuality photons. The electron behaves therefore
as a broad-band beam of real photons, whose momenta are distributed according
to the Weizs\"acker-Williams function~[\ref{WW}]
\beq
f_{\gamma}^{(e)}(y)=\frac{\aem}{2\pi}
\left[
\frac{1+(1-y)^2}{y}\log\frac{Q^2(1-y)}{m_e^2 y^2}
+2m_e^2y\left(\frac{1}{Q^2}-\frac{1-y}{m_e^2 y^2}\right)\right],
\label{wwfun}
\eeq
where $m_e$ is the electron mass and $y$ is the fraction of the electron
longitudinal momentum carried by the photon. Notice that in eq.~(\ref{wwfun})
terms which are non-singular for $m_e\to 0$ cannot be neglected in the HERA
energy range, because of the $1/y$ singularity~[\ref{WW2},\ref{FMNRWW}].
The quantity $Q$ must be taken to be the minimum between the typical 
energy scale which characterizes the production process, and the upper limit 
of the absolute value of the photon virtuality (see ref.~[\ref{FMNRWW}] 
for a detailed discussion of this point). The cross section for a 
generic electron-proton scattering process is given by 
\beq
d\sigma_{ep}(K_e,K_p)=\int_{y_{min}}^{y_{max}}
dy \,f_\gamma^{(e)}(y)\,d\sigma_{\gamma p}(yK_e,K_p),
\label{sigep}
\eeq
where $K_e$ and $K_p$ are the momenta of the incoming electron and
proton respectively, and \mbox{$0<y_{min}<y_{max}\leq 1$} are fixed by
kinematical boundary conditions or experimental cuts.
The relevant hard scattering quantity is therefore 
the cross section for photon-proton collisions, $d\sigma_{\gamma p}$.
Since the HERA experiments can measure the momentum of the
scattered electron down to very small values of the scattering angle,
it is possible, in principle, to measure directly $d\sigma_{\gamma p}$ 
for a fixed value of the photon energy (with the current electron
and proton energies, $E_e=27.5$~GeV and $E_p=820$~GeV, the accessible 
range in the photon-proton center-of-mass energy is at best 
$100$~GeV$\leq E_{cm}(\gamma p)\leq 280$~GeV).

\begin{figure}
\centerline{\epsfig{figure=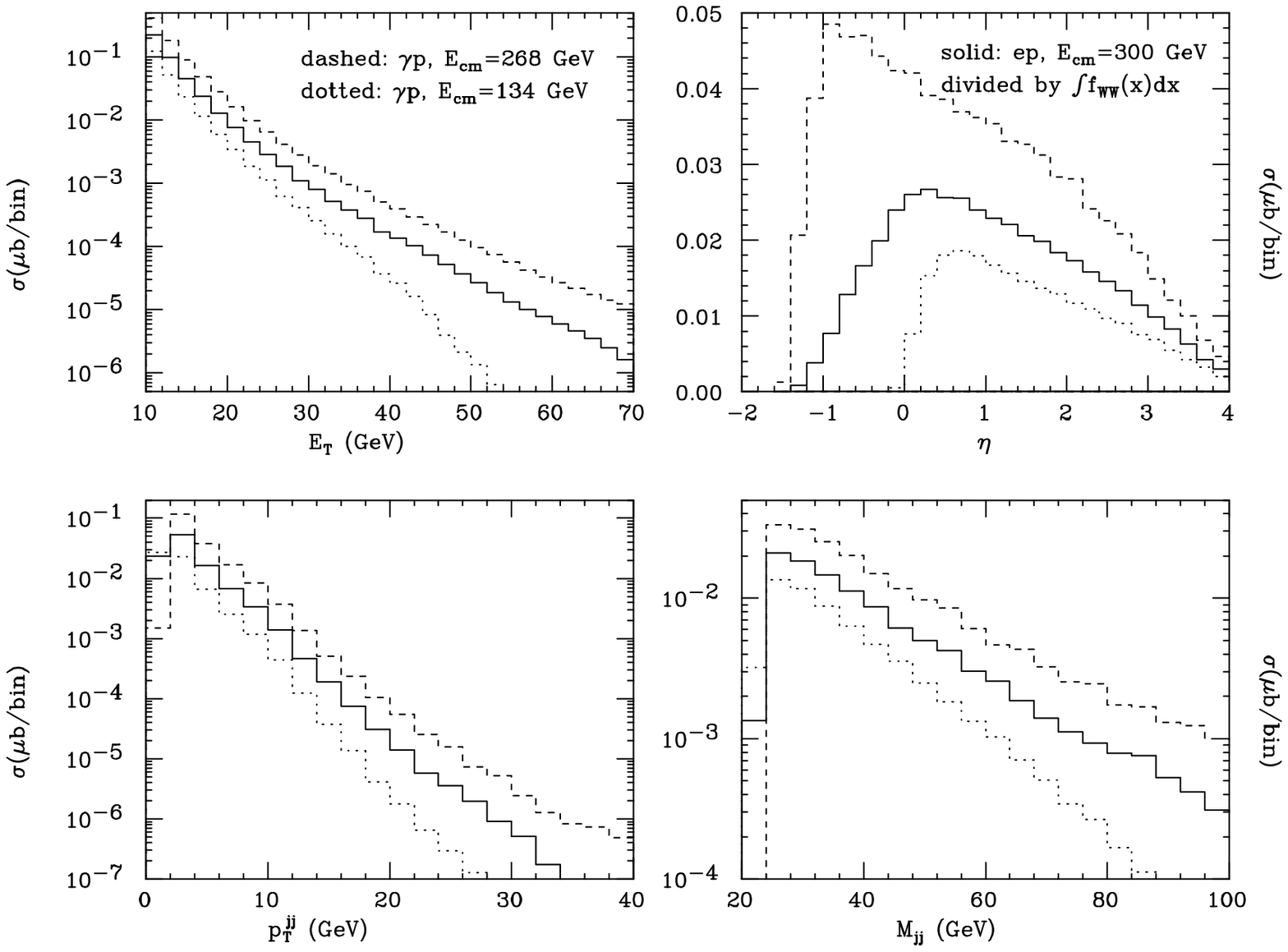,width=\textwidth,clip=}}
\ccaption{}{ \label{fig:monoch}  
Comparison between monochromatic photon-proton and electron-proton
(Weizs\"acker-Williams approximation) cross sections, for various
jet distributions.
}
\end{figure}

The possibility of measuring the photon-proton jet cross section for
several different center-of-mass energies would be an interesting
test of the scaling laws predicted by QCD. For this reason, we have
studied the energy dependence of differential jet distributions in
monochromatic photon-proton collisions, comparing them with the
same quantities obtained in electron-proton collisions in the
Weizs\"acker-Williams approximation, eq.~(\ref{sigep}). The results,
obtained at NLO in QCD, are shown in fig.~\ref{fig:monoch}.
In the two upper plots we present single-inclusive distributions
in the transverse energy $\et$ of the jet and in its pseudorapidity $\eta$
for $\et>\etcut$, $\etcut=10$~GeV. In the two lower plots, we
show the distributions in the transverse momentum, $p_{\sss T}^{jj}$, 
and invariant mass, $M_{jj}$, of the pair of the two
hardest jets in the event. In this case, we require the transverse
energies of the two observed jets to satisfy the constraints
$E_{1\sss T}>E_{1\sss T}^{cut}$, $E_{2\sss T}>E_{2\sss T}^{cut}$,
with $E_{1\sss T}^{cut}=E_{2\sss T}^{cut}=10$~GeV (the fact that the
same value of the transverse energy cut is used for both jets will 
be discussed below in detail).
Jets are defined by the cone algorithm~[\ref{conealg}] 
with opening angle $R=1$. We used the sets MRSA$^\prime$~[\ref{MRSAprime}]
and GRV-HO~[\ref{GRVPDFpho}]
for parton densities in the proton and the photon respectively.
The renormalization and factorization scales have been chosen equal
to half the transverse energy of the event; from now on, this
choice will be our default and will be denoted as $\mu_0$. 
$\Lambda_{\sss QCD}$ has been set equal
to the value suggested by the MRSA$^\prime$ parton densities, 
$\Lambda_5^{\rm \overline{MS}}=152$~MeV.
Each distribution in fig.~\ref{fig:monoch} is presented for two different 
energies of the photon-proton system, 134~GeV (dotted curves) and 268~GeV
(dashed curves), which correspond to $y=0.2$ and $y=0.8$ in eq.~(\ref{sigep})
respectively. The solid curves show the predictions for electron-proton 
collisions at $E_{cm}(ep)=300$~GeV in the Weizs\"acker-Williams
approximation, eq.~(\ref{sigep}), with $y_{min}=0.2$, $y_{max}=0.8$,
and $Q^2=4$~GeV$^2$, rescaled by a factor
\beq
\Bigg[\int_{y_{min}}^{y_{max}}f_\gamma^{(e)}(y)\, dy\Bigg]^{-1}\simeq 26.9\,.
\label{normWW}
\eeq

The difference in normalization between the curves relevant for
the two photon-proton center-of-mass energies is sizeable.
Phase-space effects are clearly visible in the tail of the
$\et$, $p_{\sss T}^{jj}$ and $M_{jj}$ distributions, where the
prediction for $E_{cm}=134$~GeV decreases faster than the power-like
behaviour typical of QCD, and in the range of negative $\eta$
(the photon is coming from the right, as in the HERA conventions).
By construction, there is no dynamical information on the hard
scattering process in the Weizs\"acker-Williams approximation,
which is simply a weighted sum of $\gamma p$ cross sections;
this is also clear from the shape of the solid curves in 
fig.~\ref{fig:monoch}.
For this reason, in this section we will mainly consider monochromatic
photon-proton collisions at 
$E_{cm}=268$~GeV. We have verified that our
conclusions apply down to the lowest center-of-mass energies
relevant at HERA.

We now turn to the problem of studying the reliability of the
perturbative expansion. To this purpose, we have considered
photon-proton collisions at 268~GeV as a representative case
and we have computed various
distributions, varying factorization and renormalization scales by a 
factor of 2 around our default choice, as customary in QCD
in order to assess the magnitude of the neglected higher-order
corrections. 
QCD predictions for photon-hadron cross sections are computed in terms of two 
distinct components: the so-called pointlike component, in which the photon
directly interacts with partons in the hadron, and the hadronic component,
in which the photon fluctuates into hadronic states, which subsequently
interact with the hadron. This distinction is well defined at leading order,
but becomes arbitrary when higher order terms in the perturbative expansion
are included (we will thoroughly discuss this problem in the following). 
Explicitly, the cross section is written as
\beq
d\sigma_{\gamma p}(K_\gamma,K_p)=d\sigma^{\rm point}_{\gamma p}(K_\gamma,K_p)
+d\sigma^{\rm hadr}_{\gamma p}(K_\gamma,K_p)
\label{sigmagp}
\eeq
where
\beqn
d\sigma^{\rm point}_{\gamma p}(K_\gamma,K_p)&=&\sum_j\int dx
f^{(p)}_j(x,\muhp)
d\hat{\sigma}_{\gamma j}(K_\gamma,xK_p,\as(\murp),\murp,\muhp,\mug),
\label{pointcomp}
\\
d\sigma^{\rm hadr}_{\gamma p}(K_\gamma,K_p)
&=&\sum_{ij}\int dx dy
f^{(\gamma)}_i(x,\mug) f^{(p)}_j(y,\muhs)
\nonumber \\*&&\phantom{\sum_{ij}\int dx}\times
d\hat{\sigma}_{ij}(xK_\gamma,yK_p,
\as(\murs),\murs,\muhs,\mug)\,,
\label{hadrcomp}
\eeqn
and $f^{(p)}_i (f^{(\gamma)}_i)$ are the distribution functions
of parton $i$ inside the proton (photon). $d\hat{\sigma}_{\gamma j}$ and
$d\hat{\sigma}_{ij}$ are the subtracted short-distance partonic cross
sections, which are finite at any order in perturbative QCD. 
The dependence of 
$d\sigma^{\rm point}_{\gamma p}$ on $\murp,\muhp$, and of
$d\sigma^{\rm hadr}_{\gamma p}$ on $\murs,\muhs$,
cancels in each component separately, up to NNLO terms. On the other hand,
the dependence on $\mug$ in $d\sigma^{\rm point}_{\gamma p}$ is 
compensated by the analogous dependence in $d\sigma^{\rm hadr}_{\gamma p}$ 
(see for example ref.~[\ref{FMNRPF}] for a detailed discussion on this point).
In the language of Altarelli-Parisi equations, the collinear singularities
of the photon leg in the pointlike component are re-absorbed (at a scale
$\mug$) in the parton densities of the photon entering the hadronic
component. 

We begin by studying single-inclusive quantities and two-jet
observables, defined with $\etonecut\neq\ettwocut$. We set all
the scales in eqs.~(\ref{pointcomp}) and~(\ref{hadrcomp}) to the
same value, which we denote by $\mu$, and we study the dependence
of the physical quantities upon $\mu$. In this case, this procedure
gives a good estimate of the full uncertainty affecting the results,
which in principle should be obtained by considering independent
variations of all the scales.
\begin{figure}
\centerline{
   \epsfig{figure=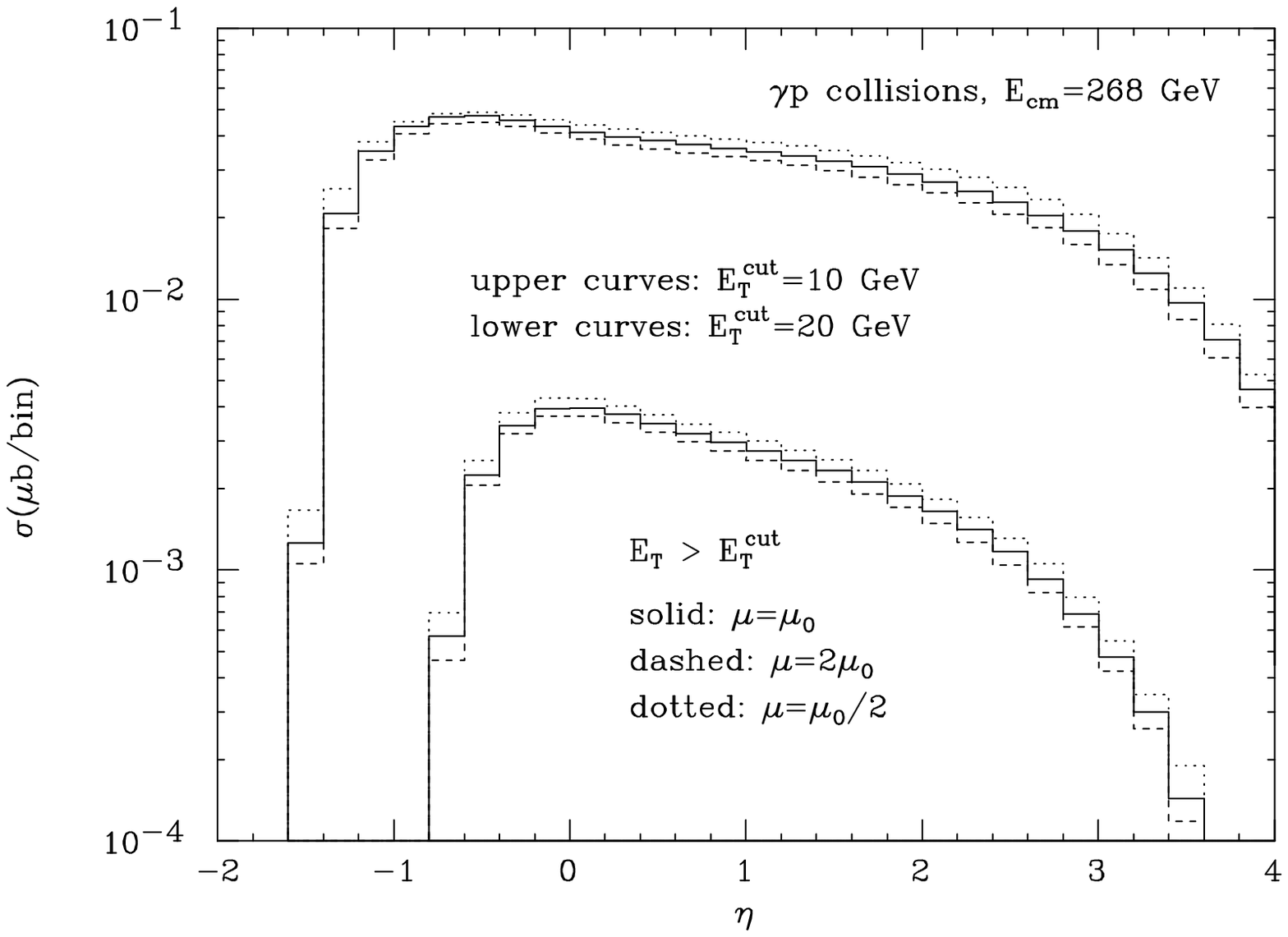,width=0.48\textwidth,clip=}
   \hfill
   \epsfig{figure=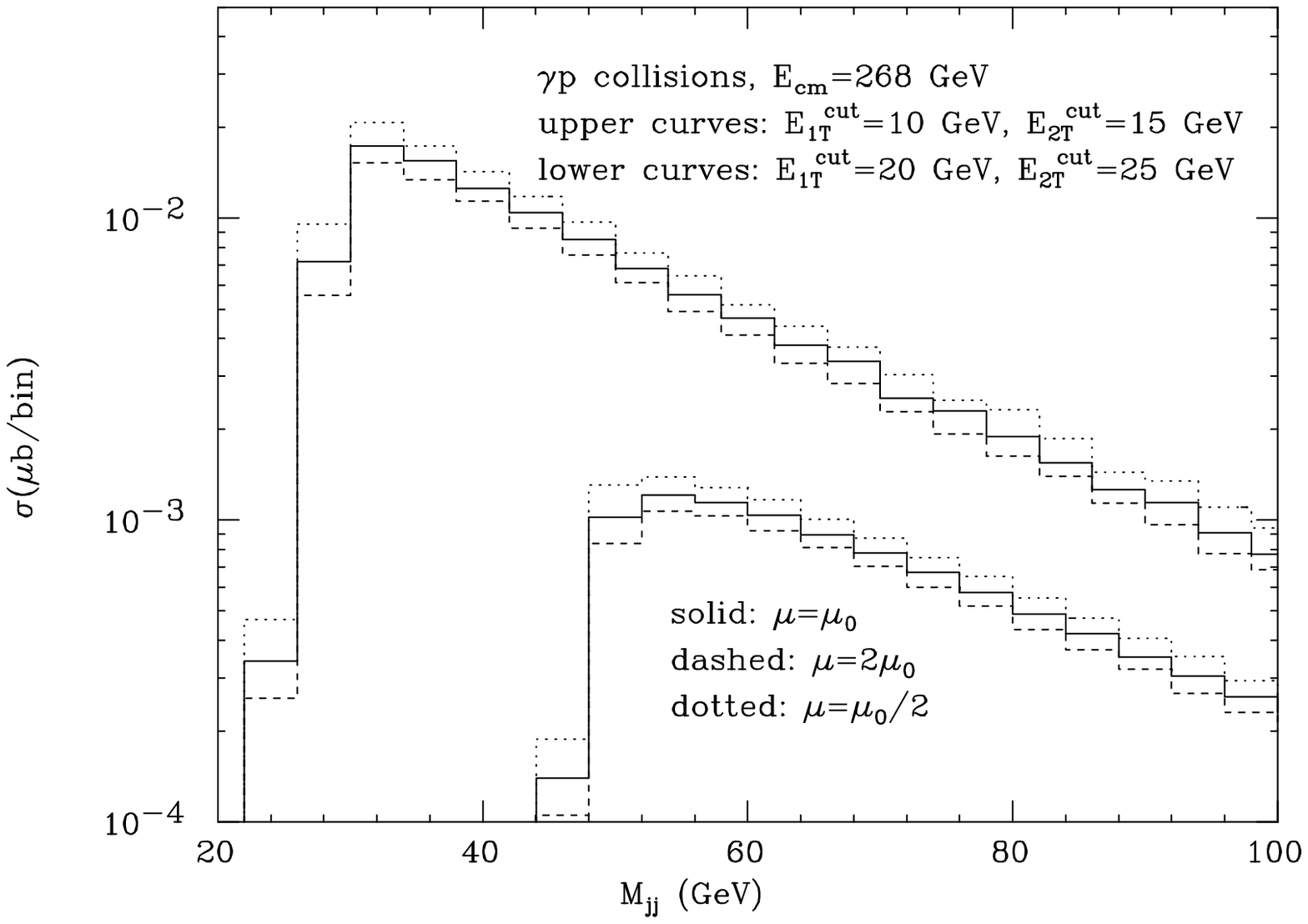,width=0.48\textwidth,clip=} }
\ccaption{}{ \label{fig:psc1_4_feta_mjj}
Scale dependence of the single-inclusive jet pseudorapidity and two-jet
invariant mass distributions, for different values of the minimum allowed
transverse energy of the observed jets.
}
\end{figure}                                                              
\begin{figure}
\centerline{\epsfig{figure=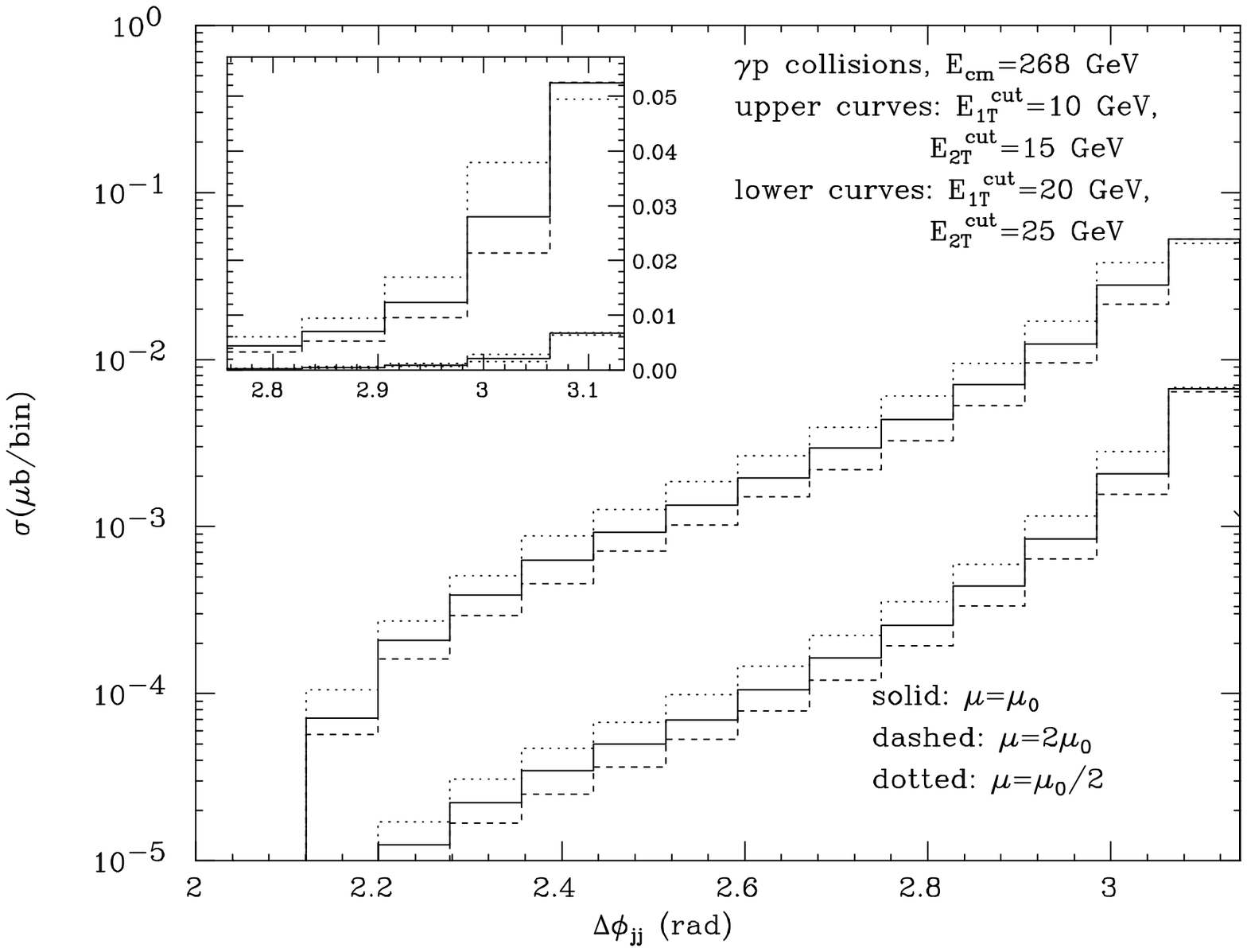,width=0.7\textwidth,clip=}}
\ccaption{}{ \label{fig:psc1_4_fphi_a}  
Scale dependence of the azimuthal distance distribution in two-jet events, 
for different values of the minimum allowed transverse energy of the 
two hardest jets ($\etonecut\neq\ettwocut$).
}
\end{figure}                                                              
In the left plot of fig.~\ref{fig:psc1_4_feta_mjj} we present
the distribution in the pseudorapidity of jets with $\et>E_{\sss T}^{cut}$,
for $E_{\sss T}^{cut}=10$~GeV and 20~GeV. In the right plot of the same figure
we show the distribution in the invariant mass of the two hardest jets,
for $E_{1\sss T}^{cut}=10$~GeV, $E_{2\sss T}^{cut}=15$~GeV (upper curves)
and $E_{1\sss T}^{cut}=20$~GeV, $E_{2\sss T}^{cut}=25$~GeV (lower curves).
The jets are defined by the cone algorithm with $R=1$.
The scale dependence is moderate, and slightly reduces with 
increasing transverse energy cuts. We checked that other 
kinematical quantities, like the $\et$ of the jet, the pseudorapidity
of the pair, the difference in rapidity of the pair, show a similar scale
dependence. From the left-hand side of fig.~\ref{fig:psc1_4_feta_mjj},
we see that the scale dependence of the pseudorapidity distribution is 
slightly stronger for $\eta$ values which are far from the central region.
This is what we expect, since large $\eta$ values correspond to 
small transverse energies, and therefore to less reliable QCD predictions.

We also computed the same quantities for a cone with an
opening angle of $R=0.7$. In this case, the scale dependence of both 
the pseudorapidity and the invariant mass distributions is slightly reduced.

The situation is different for the azimuthal correlation $\Delta\phi_{jj}$ 
between the two jets with largest transverse momenta, shown in
fig.~\ref{fig:psc1_4_fphi_a}. For $\Delta\phi_{jj}<\pi$ the two sets of 
curves corresponding to the different choices of transverse momentum cuts
show the same scale dependence, larger than in the case of 
pseudorapidity and invariant mass distributions of 
fig.~\ref{fig:psc1_4_feta_mjj}. This is due to the fact that the 
$\Delta\phi_{jj}$ correlation is a pure NLO effect in this region.
The scale dependence reduces for $\Delta\phi_{jj}\simeq\pi$, as can
be also seen from the small inserted figure.

The $\Delta\phi_{jj}$ correlation computed with a cone of
$R=0.7$ shows the same scale dependence as in the case $R=1$ for
$\Delta\phi_{jj}<\pi$, and a stronger scale dependence for
$\Delta\phi_{jj}\simeq\pi$.

\begin{figure}
\centerline{\epsfig{figure=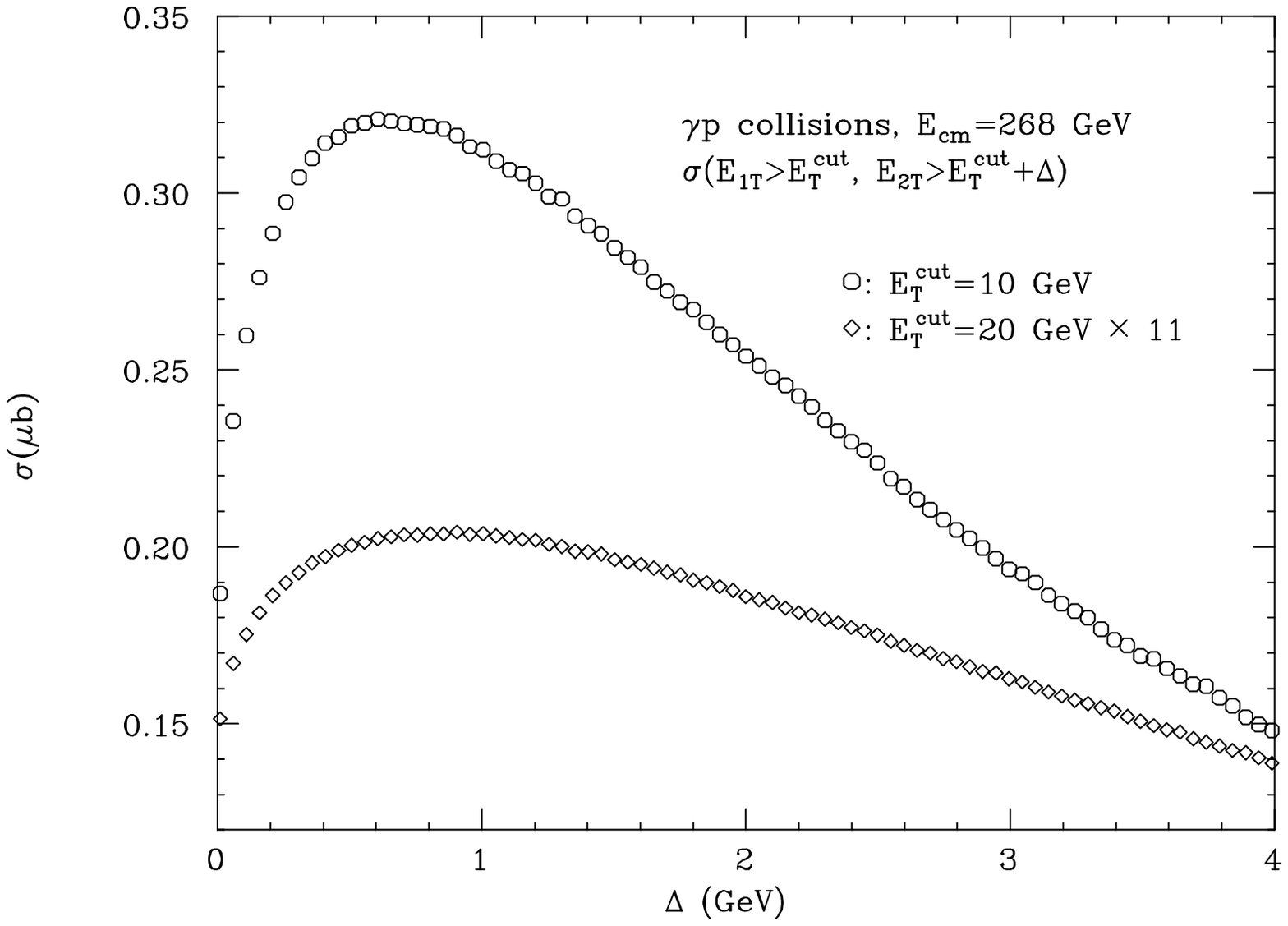,width=0.7\textwidth,clip=}}
\ccaption{}{ \label{fig:sigmatot}  
Inclusive two-jet total cross section for $\etone>\etcut$
and \mbox{$\ettwo>\etcut+\Delta$}, as a function of $\Delta$, for two
different values of $\etcut$.
}
\end{figure}

We now consider the problem of computing two-jet inclusive quantities for 
$\etonecut=\ettwocut\equiv\etcut$. As far as infrared 
safeness is concerned, there is nothing special in this choice. The
cross section is well-defined and finite at any order in perturbation
theory. On the other hand, there are quantities which at next-to-leading
order display a pathological behaviour. This can be seen very easily
by studying the inclusive two-jet total cross section 
\beq
\sigma_2(\Delta)=\sigma(E_{1\sss T}>\etcut,E_{2\sss T}>\etcut+\Delta)
\eeq
for $\Delta\to 0$. The next-to-leading order QCD results are shown 
in fig.~\ref{fig:sigmatot}, where we have chosen $\etcut=10$~GeV
and $\etcut=20$~GeV (the curve for $\etcut=20$~GeV has been rescaled by
a factor of 11 in order to make both curves visible on the
same plot).
Notice that the value of $\sigma_2(\Delta)$ for $\Delta=0$ is finite,
as expected for an infrared-safe quantity. Observe also, on the 
other hand, that $\sigma_2(\Delta)$ has an infinite slope in $\Delta=0$.
This fact can be understood in the following way. We consider
the real emission contribution when one of the emitted
partons is quasi-collinear
to one of the initial state partons (at NLO in QCD this implies that
the two jets are identified with the two other partons). The leading
collinear singularity of this contribution is given by
\beq
\sigma^{(r)}= \int d^2p_{1\sss T}\,\theta(E_{1\sss T}-\etcut)
\int d^2p_{2\sss T}\,\theta(E_{2\sss T}-\etcut-\Delta)\,
\frac{1}{\abs{\vec{p}_{1\sss T}+\vec{p}_{2\sss T}}^2+\delta},
\eeq
where $\delta$ acts as a collinear cutoff. The integral can be easily 
computed, and we get
\beq
\sigma^{(r)} = A(\Delta,\delta)+B\log\delta
-C \cdot (\Delta+\delta)\log(\Delta+\delta),
\label{cervi}
\eeq
where both $A(\Delta,\delta)$ and its first derivative with respect to
$\Delta$ are regular in $\Delta=0$ for any $\delta$, including $\delta=0$,
and $C$ is a positive coefficient.
The term $B\log\delta$ is the genuine collinear singularity, and it is
cancelled by the corresponding singular terms in the two-body contribution.
For fixed $\Delta\neq 0$, one can safely take $\delta=0$ in the last term. 
For $\Delta\to 0$ this term vanishes, but its first derivative with respect 
to $\Delta$ diverges, and this is the origin of the behaviour observed
in the full calculation, shown in fig.~\ref{fig:sigmatot}. Alternatively,
one could have started with $\Delta=0$, as we did for the distributions
shown in fig.~\ref{fig:monoch}. In this case, the last term in 
eq.~(\ref{cervi}) becomes $\delta\log\delta$, which vanishes
for $\delta\to 0$, although less rapidly than terms linear in $\delta$.
Because of this extra $\delta$ dependence, the $\Delta=0$ case is
a reason of concern~[\ref{HO}] when the calculation is performed with 
a technique which requires keeping $\delta\neq 0$, as for example 
the slicing method. We remark that in the subtraction method the
introduction of the non-physical parameter $\delta$ is avoided,
and therefore the $\Delta=0$ case does not require any special care.

One striking feature of fig.~\ref{fig:sigmatot} is that
$\sigma_2(\Delta)$ is not a monotonically decreasing function 
for increasing $\Delta$, as one might expect on the basis of simple
phase-space considerations. This behaviour is also present is
our simple model, eq.~(\ref{cervi}), because of the negative sign in 
front of last term. This effect is induced by the truncation
of the perturbative series at NLO. Roughly speaking, at this order there is
not enough emission of soft real gluons to compensate for the large negative 
contribution of the soft-virtual terms. We therefore expect quantities like 
$\sigma_2(\Delta)$ to be poorly predicted by NLO QCD for $\Delta=0$.
However, the mismatch between virtual and real
contributions is effective only in some special regions of the phase space; 
typical examples are the threshold of the invariant mass of the pair and the
region around $\Delta\phi_{jj}=\pi$ in the azimuthal distance correlation
of the two hardest jets. It is well known that fixed-order perturbation
theory is not reliable in these regions, and that resummation of all-order 
contribution is needed. Despite this fact, the cross section is expected
to be well-behaved in the remaining part of the phase space. 

\begin{figure}
\centerline{\epsfig{figure=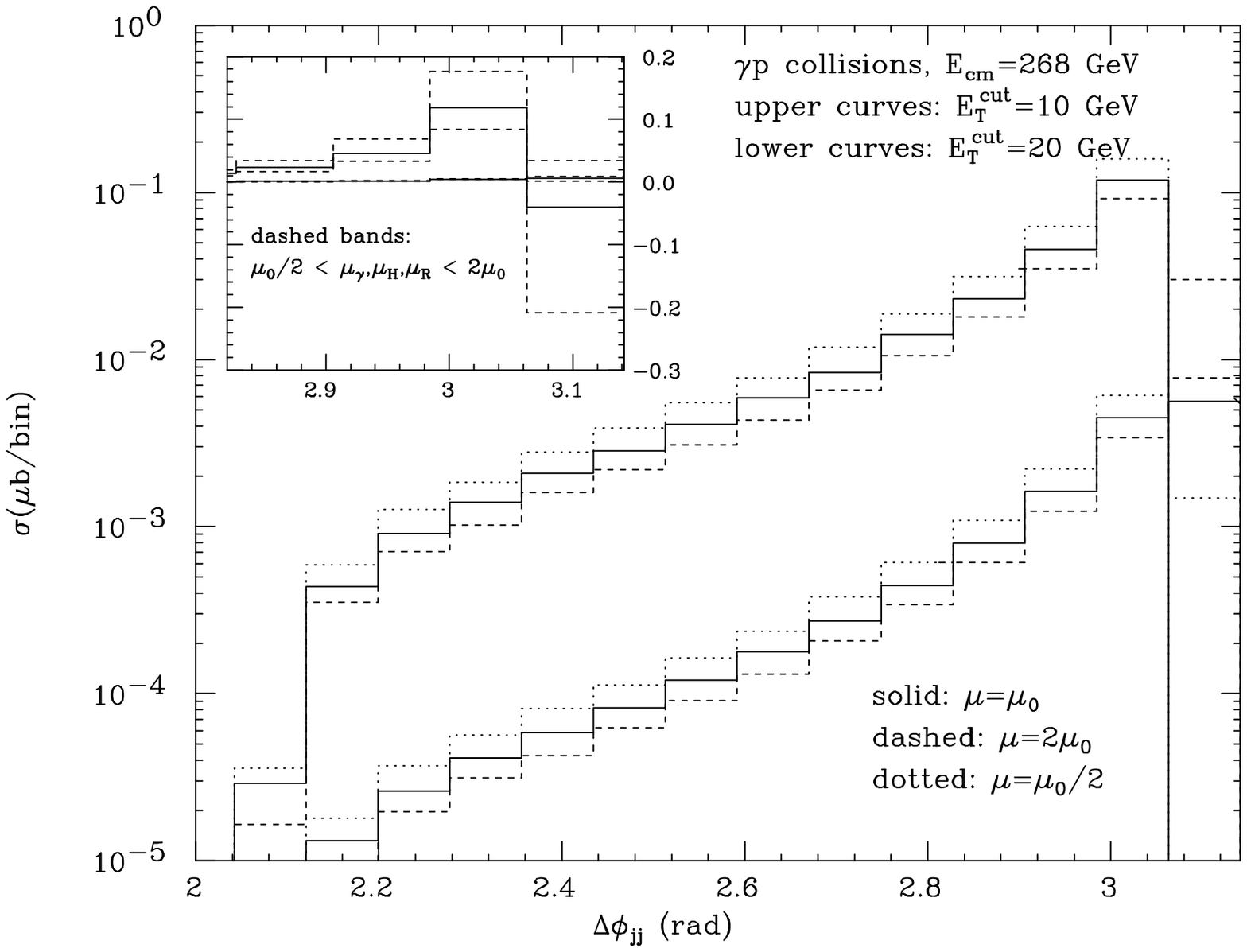,width=0.7\textwidth,clip=}}
\ccaption{}{ \label{fig:psc1_4_fphi}  
Scale dependence of the azimuthal distance distribution in two-jet events, 
for different values of the minimum allowed transverse energy of the 
two hardest jets ($\etonecut=\ettwocut=\etcut$).
}
\end{figure}

In the following, we will show that two-jet inclusive distributions can 
be safely studied even in the case of equal transverse energy cuts 
on both jets. To address this issue, we performed a more careful study 
of the scale dependence of the results. 
In particular, we varied independently $\mug$ and $\mur=\muh$,
where $\mur=\murp$ and $\muh=\muhp$ when we consider the pointlike
component, eq.~(\ref{pointcomp}), while $\mur=\murs$ and $\muh=\muhs$
when we consider the hadronic component, eq.~(\ref{hadrcomp}).
We also verified that independent variations of $\mur$ and $\muh$
would not modify our conclusions. We begin with the $\Delta\phi_{jj}$
correlation, shown in fig.~\ref{fig:psc1_4_fphi}.
The region $\Delta\phi_{jj}<\pi$ does not pose 
any problem, being a pure NLO effect. For this reason, the curves
were calculated again with $\mug=\mur=\muh$. As in the case 
$\etonecut\neq\ettwocut$, no improvement in the scale dependence is 
seen when transverse energy cuts are increased. The small inserted figure 
shows the region $\Delta\phi_{jj}\simeq\pi$; in this case, the curves 
have been evaluated by varying the scales independently.
In the bin around $\Delta\phi_{jj}=\pi$ the curves corresponding to 
$\etcut=10$~GeV show a very strong scale dependence, thus signalling 
a failure of the perturbative expansion in this region. Notice also
that most of the uncertainty band in the last bin gives 
negative cross sections. By considering independent scale 
variations, this band increases of about 25\% with respect to the case
when all the scales are set to the same value.
When $E_{\sss T}^{cut}$ is increased to 20~GeV, this effect is more
moderate, but the scale dependence is still larger than in 
the region $\Delta\phi_{jj}<\pi$ (this is consistent with 
fig.~\ref{fig:sigmatot}, where we see that the total cross section
decreases faster for $\Delta\to 0$ in the case of smaller transverse
momentum cuts, thus indicating that in this case soft gluon emission
is more important). We expect that an all-order 
resummation would have a moderate impact in this case.
Of course, the above considerations are partially dependent on
the chosen bin size: enlarging the bin around $\Delta\phi_{jj}=\pi$
enough, the perturbative prediction becomes reliable for any choice
of $\etcut$.

\begin{figure}
\centerline{
   \epsfig{figure=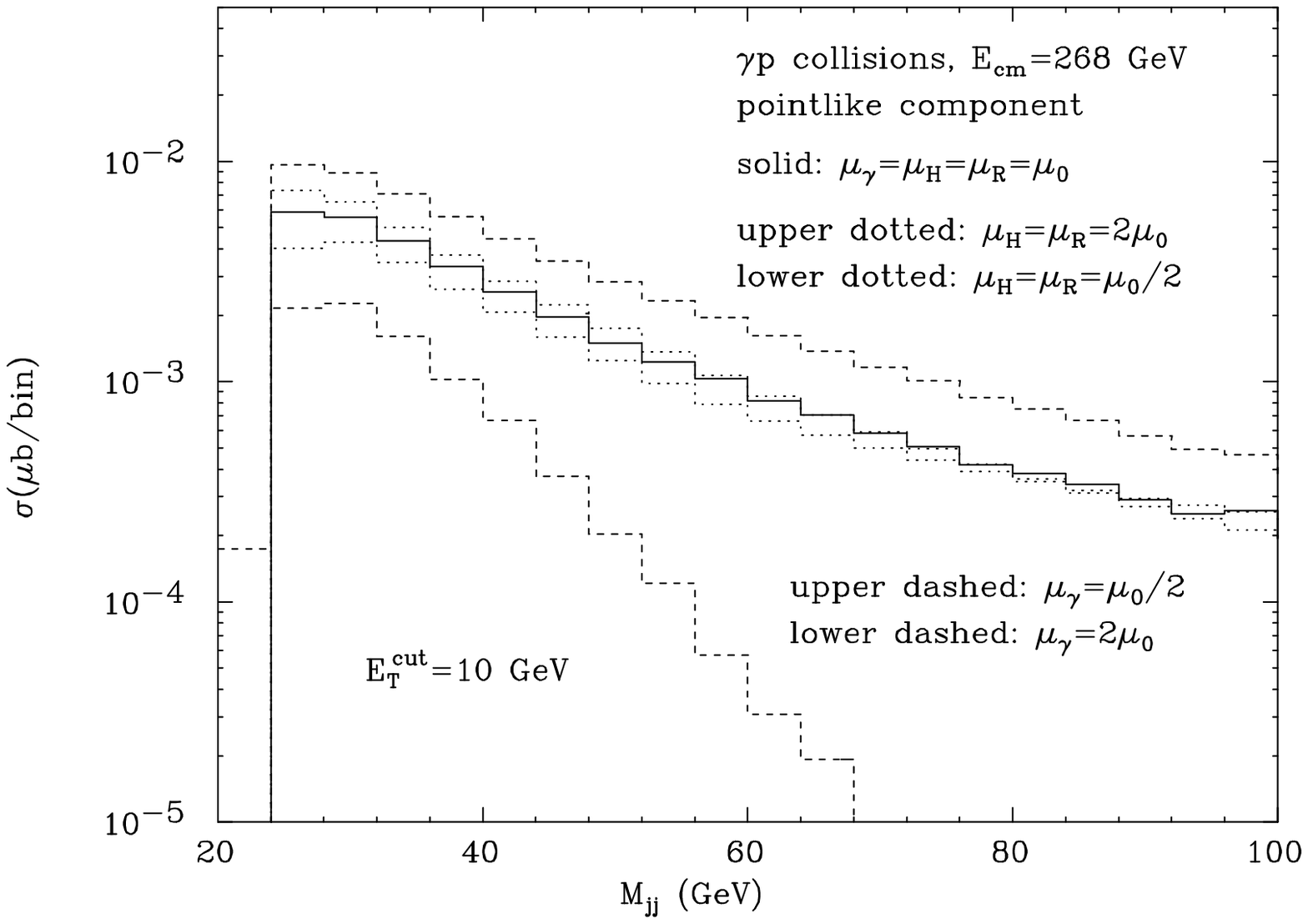,width=0.48\textwidth,clip=}
   \hfill
   \epsfig{figure=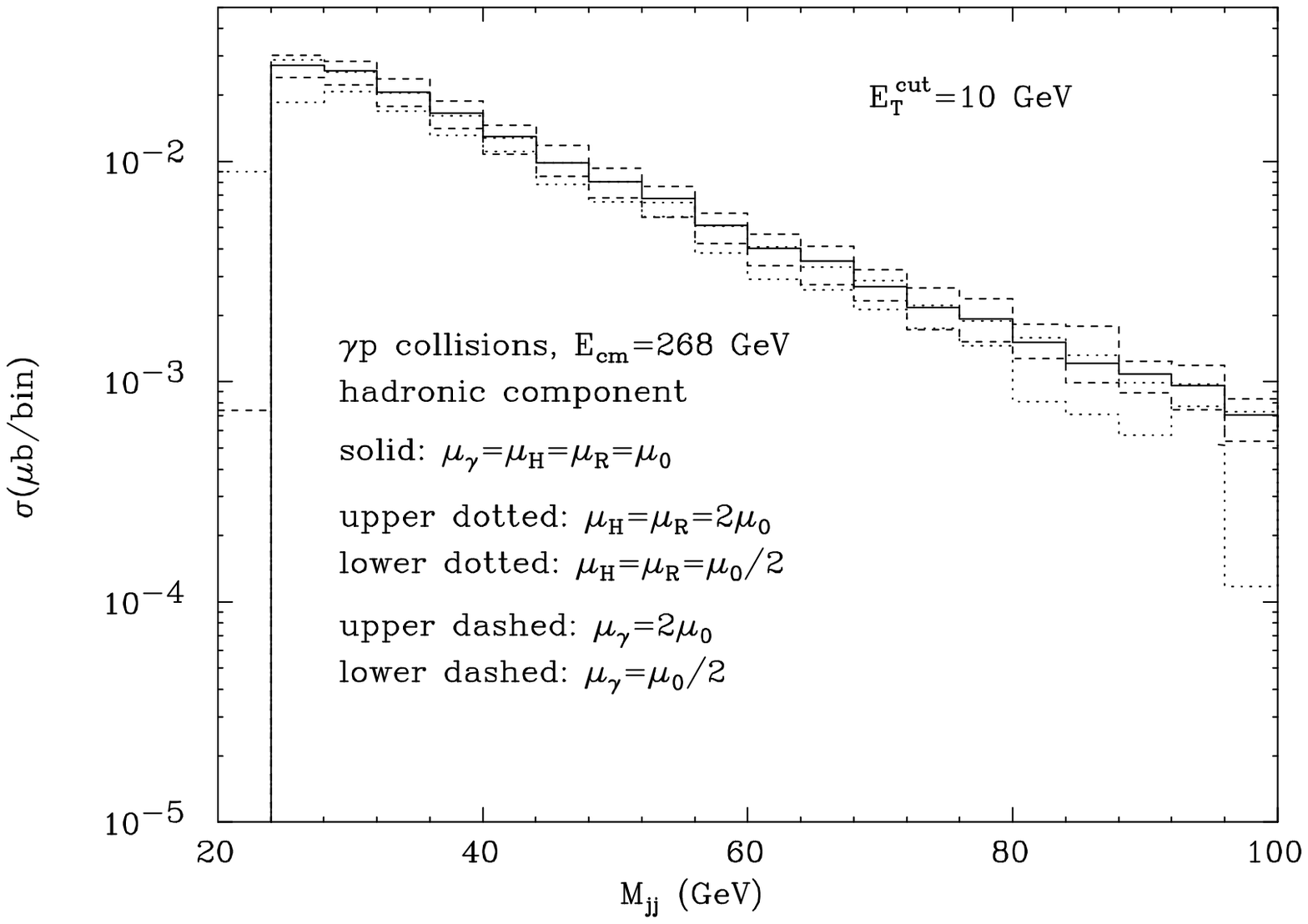,width=0.48\textwidth,clip=} }
\ccaption{}{ \label{fig:psc1_mjj3}
Scale dependence of the invariant mass distribution in two-jet events.
The pointlike (left) and hadronic (right) components are separately
shown ($\etonecut=\ettwocut=\etcut$).
}
\end{figure}

\begin{figure}
\centerline{\epsfig{figure=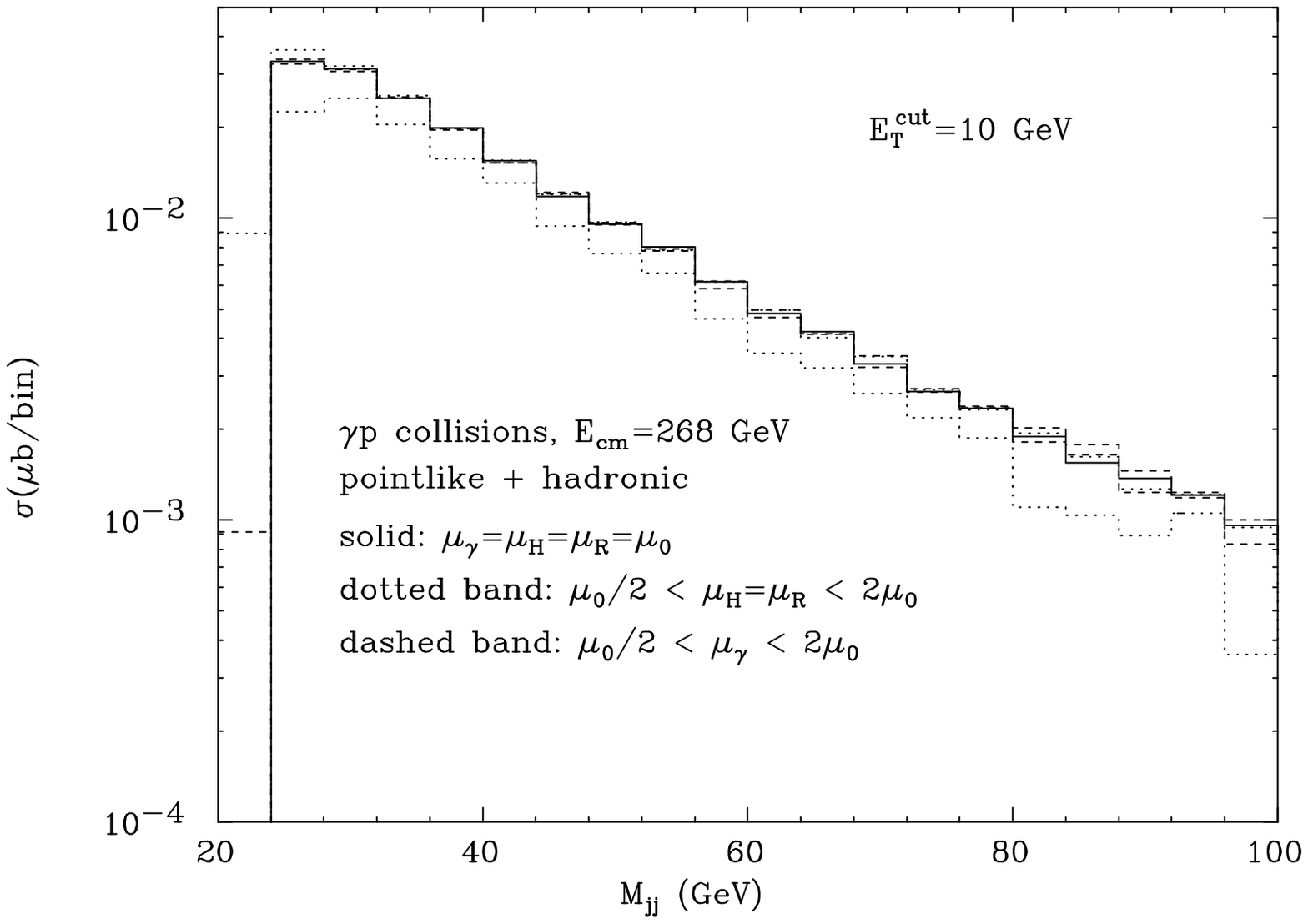,width=0.7\textwidth,clip=}}
\ccaption{}{ \label{fig:psc1_fmjj3}  
Scale dependence of the invariant mass distribution in two-jet events
($\etonecut=\ettwocut=\etcut$).
}
\end{figure}

We now turn to the distribution of the invariant mass of the pair.
This distribution is non-trivial already at leading order, and therefore,
in principle, the effect of soft gluon emission is not confined in
some particular region of the phase space. To clearly show the effect of an
independent variation of the scales, we separately present in 
fig.~\ref{fig:psc1_mjj3} the pointlike and the hadronic contributions,
for the choice $\etonecut=\ettwocut=10$~GeV.
The solid curves are the defaults, the dotted (dashed) curves are 
obtained by varying $\mur=\muh$ ($\mug$). We see that the pointlike
and the hadronic component display opposite behaviours with respect
to the scale $\mug$ (for $\mug=2\muo$, the curve is larger than
the default one for the hadronic component, and smaller for
the pointlike component). This shows the cancellation
mechanism between eqs.~(\ref{pointcomp}) and~(\ref{hadrcomp}).
In particular, the fact that the hadronic component is larger than
the pointlike one implies that a moderate $\mug$ dependence in the
hadronic contribution induces a huge $\mug$ dependence in the 
pointlike contribution, as can be seen from fig.~\ref{fig:psc1_mjj3}.
This does not mean that the prediction of the theory is not reliable: 
in fact, the only measurable quantity is the sum of the two components,
shown in fig.~\ref{fig:psc1_fmjj3}. From that figure, we see that
the compensation between the $\mug$ dependence of the two components
is remarkable, and the residual dependence is rather small.
On the other hand, there is no cancellation mechanism in the case
of the $\mur$ and $\muh$ dependence, since these quantities 
are different in the pointlike and hadronic components (see
eqs.~(\ref{pointcomp}) and~(\ref{hadrcomp})). The corresponding
uncertainties must be summed incoherently. Nevertheless, these
uncertainties are small, therefore resulting in a small uncertainty
on the physical result, fig.~\ref{fig:psc1_fmjj3}.

This discussion does not apply to the region where
$M_{jj}$ is close to the threshold. From fig.~\ref{fig:psc1_fmjj3}
we see that the bin around $M_{jj}=20$~GeV shows a large scale
dependence, and that some of the curves are negative. In this region,
the perturbative expansion therefore breaks down, like in the case
$\Delta\phi_{jj}\simeq\pi$ we discussed above.

We conclude that NLO QCD provides reasonably stable predictions
for jet production at HERA. Single-inclusive variables are very
well described. In the case of two-jet inclusive quantities, we
have to distinguish the cases $\etonecut\neq\ettwocut$ and
$\etonecut=\ettwocut$. In the former case, the perturbative
result is rather accurate, and gives reliable predictions in the
full range of any quantity (this assumes that $\etonecut$ and 
$\ettwocut$ are not too close to each other. By inspection of
fig.~\ref{fig:sigmatot}, it is safe to take
$\abs{\etonecut -\ettwocut}>2$~GeV). In the latter case, there are 
regions of the phase space which are particularly sensitive to soft gluon
emission, and therefore require an all-order resummation.
In these regions ($\Delta\phi_{jj}\simeq\pi$, $p_{\sss T}^{jj}\simeq 0$,
$M_{jj}$ close to the threshold) the fixed-order QCD results 
are not reliable, and no significant comparison can be made with
experimental results (which, on the other hand, can be safely
obtained, since $\etonecut=\ettwocut$ does not imply infrared
non-safeness). Elsewhere, two-jet correlations can be
predicted to a level of accuracy comparable to the one obtained
in the case $\etonecut\neq\ettwocut$

As discussed above (see in particular fig.~\ref{fig:psc1_mjj3})
the pointlike and hadronic components are strongly correlated in
perturbative QCD beyond leading order. This issue is interesting 
in the light of the fact that some analyses of experimental data 
are performed with specific kinematical cuts, imposed in order to 
measure physical quantities which are supposed to be dominated by
the pointlike or the hadronic component of the cross section
for any reasonable scale choice. In jet physics, such cuts are usually
defined in terms of the variable
\beq
x_\gamma=\frac{E_{\sss 1T} e^{-\eta_1}+E_{\sss 2T} e^{-\eta_2}}{2E_\gamma},
\eeq
where $E_{\sss iT}$ and $\eta_i$ ($i=1,2$) are the transverse energies
and pseudorapidities of the two hardest jets in the event. For two-jet
production at leading order, $x_\gamma$ is equal to 1 for the pointlike 
component (in the hadronic component $x_\gamma$ is the fraction of 
the photon longitudinal momentum carried by the interacting parton). At 
higher orders, this identification no longer holds, but one might expect 
that the high-$x_\gamma$ region is dominated by the pointlike component,
and the low-$x_\gamma$ region by the hadronic one.
Following this criterion, in some analyses the pointlike (hadronic) 
component is {\it operationally} defined as the measured cross section 
for $x_\gamma$ larger (smaller) than some fixed value, usually 0.75. 
With such a definition, one might for example hope to pose 
direct constraints on the parton densities in the photon, or to
get information on the details of the underlying parton dynamics,
using jet data. 

\begin{figure}
\centerline{
   \epsfig{figure=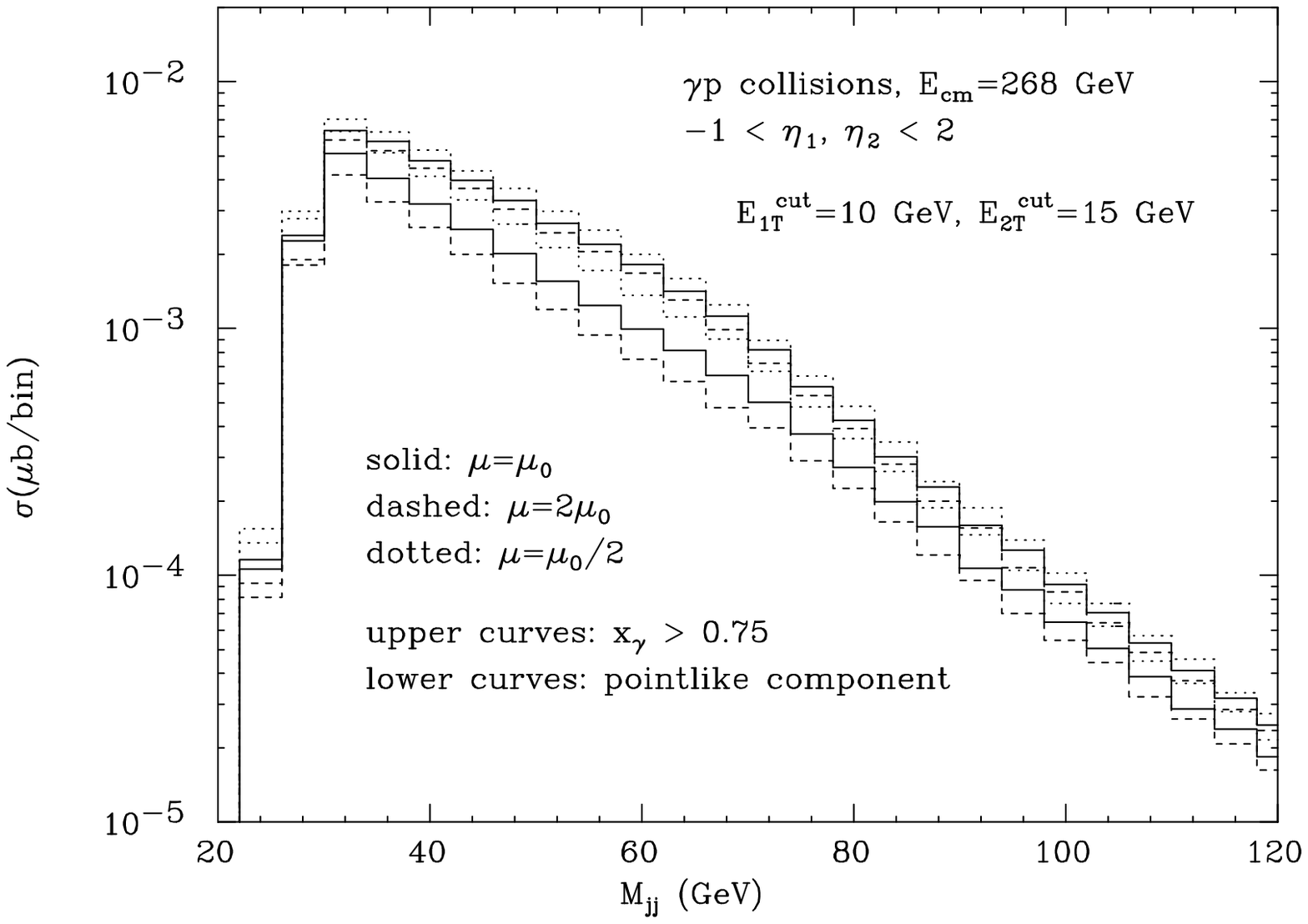,width=0.48\textwidth,clip=}
   \hfill
   \epsfig{figure=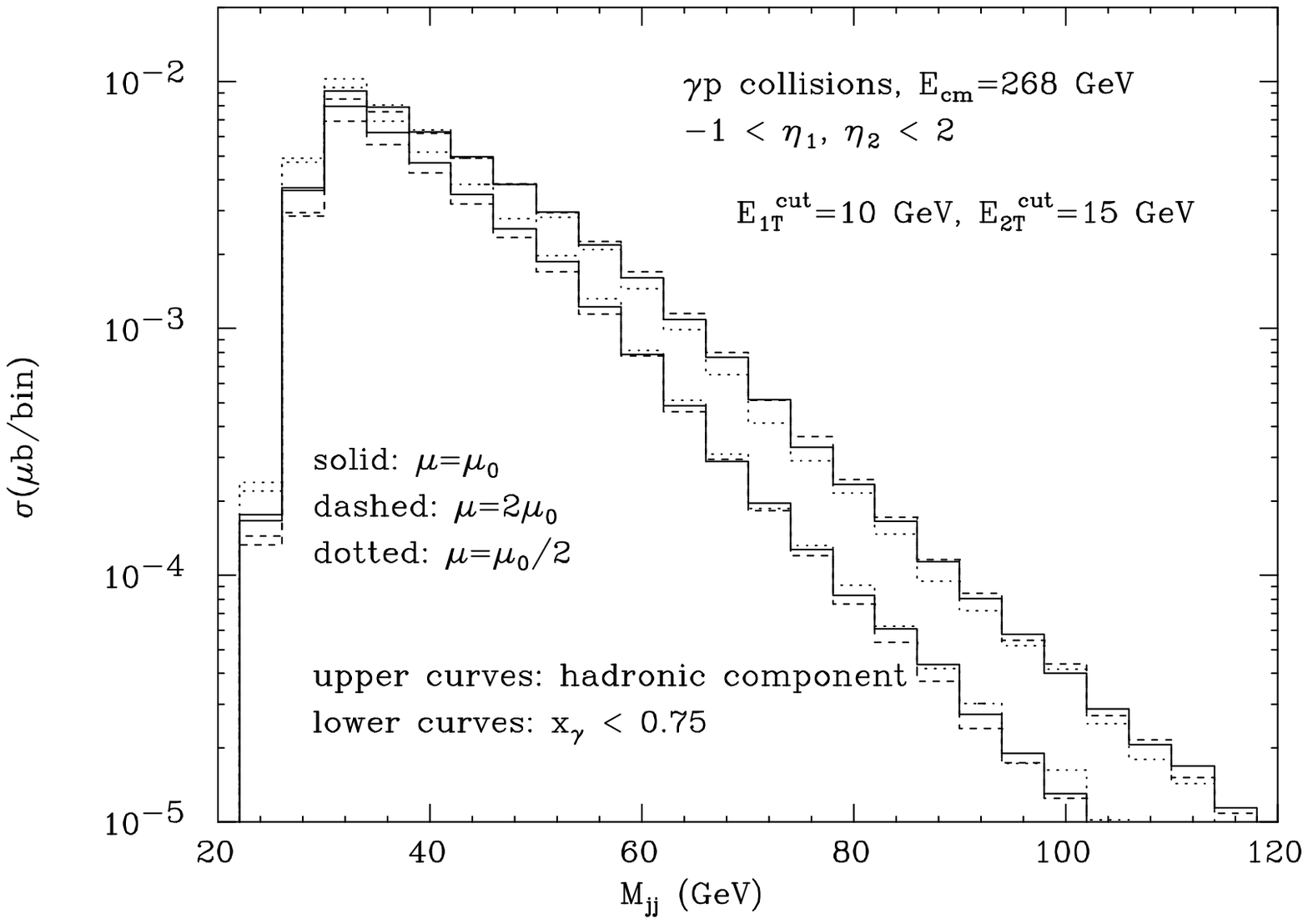,width=0.48\textwidth,clip=} }
\ccaption{}{\label{fig:psc1_mjj}
Scale dependence of the invariant mass distribution, for  
$x_{\gamma}>0.75$ (left) and $x_{\gamma}<0.75$ (right), compared with 
the pointlike and hadronic component respectively. 
}
\end{figure}                                                              
In order to assess the effectiveness of this procedure, we have computed 
the pair invariant mass distribution imposing the cuts $x_\gamma>0.75$ or
$x_\gamma<0.75$. We considered again $\gamma p$ collisions at a 
center-of-mass energy of 268~GeV, defining the jet by the cone 
algorithm with $R=1$. In fig.~\ref{fig:psc1_mjj} we present our
results for the choice $\etonecut=10$~GeV, $\etonecut=15$~GeV.
In the figure on the left (right), we show our prediction for 
the curve with $x_\gamma>0.75$ ($x_\gamma<0.75$), together with
the pointlike (hadronic) component. As before in the case 
$\etonecut\neq\ettwocut$, the scale dependence has been studied by
setting all the scales to the same value. The cuts \mbox{$-1<\eta_1,\eta_2<2$}
have been imposed on the pseudorapidity of 
the two observed jets in order to simulate a realistic geometrical
acceptance. We observe that the hadronic component (see 
fig.~\ref{fig:psc1_mjj} right) is sizeably larger than the distribution
for $x_\gamma<0.75$. This implies that in the region $x_\gamma<0.75$
the pointlike and the hadronic contributions mix significantly
for any scale choice. The same pattern is displayed by the 
pointlike component and the distribution for $x_\gamma>0.75$
(fig.~\ref{fig:psc1_mjj} left); in this case the scale 
dependence of the pointlike component is somewhat larger.
We also computed the distributions of fig.~\ref{fig:psc1_mjj}
in the case $\etonecut=\ettwocut=10$~GeV. We observe the same 
behaviour; the difference between the hadronic (pointlike) component
and the curve for $x_\gamma<0.75$ ($x_\gamma>0.75$) is even larger than
that shown in fig.~\ref{fig:psc1_mjj}.

\begin{figure}
\centerline{
   \epsfig{figure=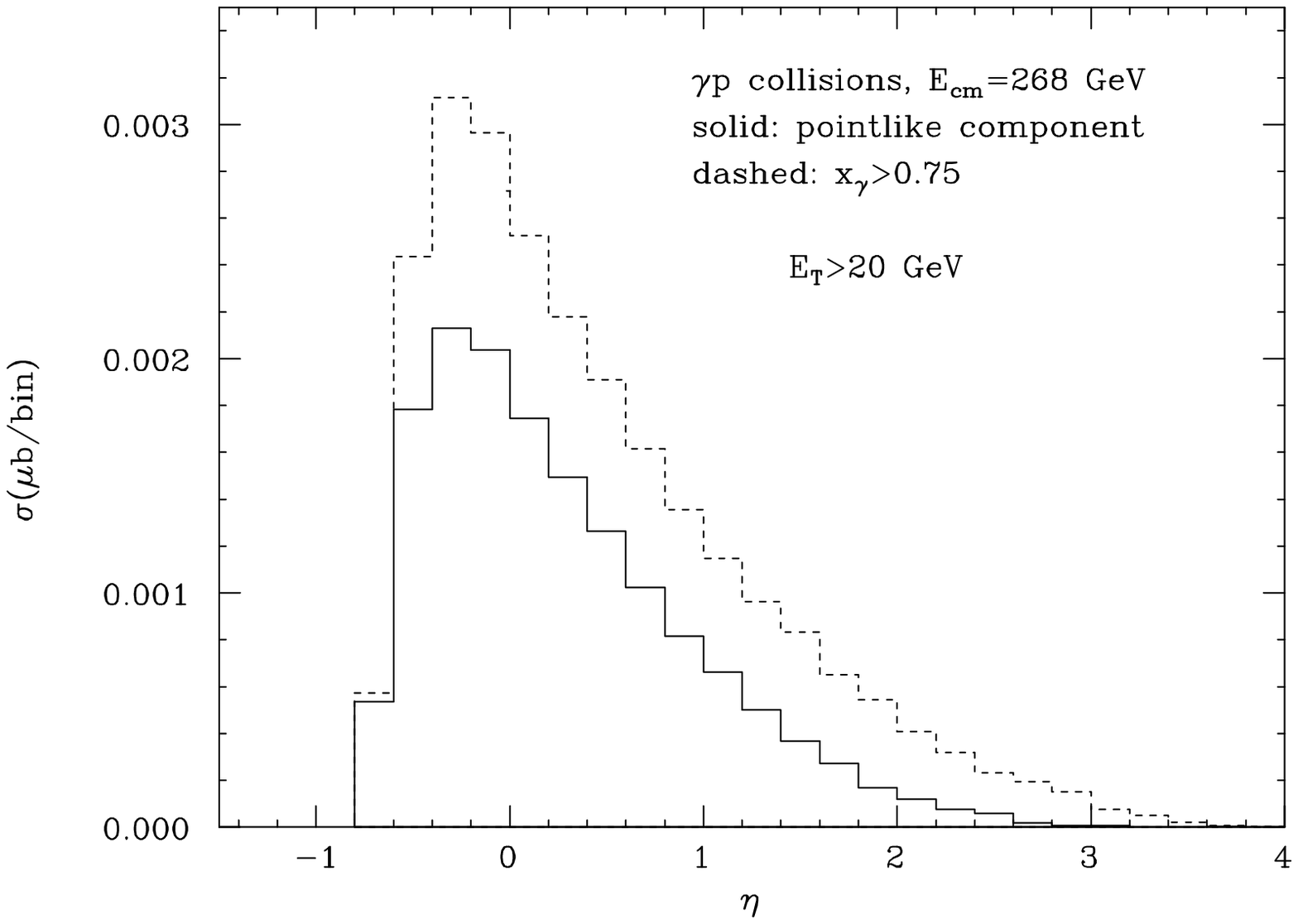,width=0.48\textwidth,clip=}
   \hfill
   \epsfig{figure=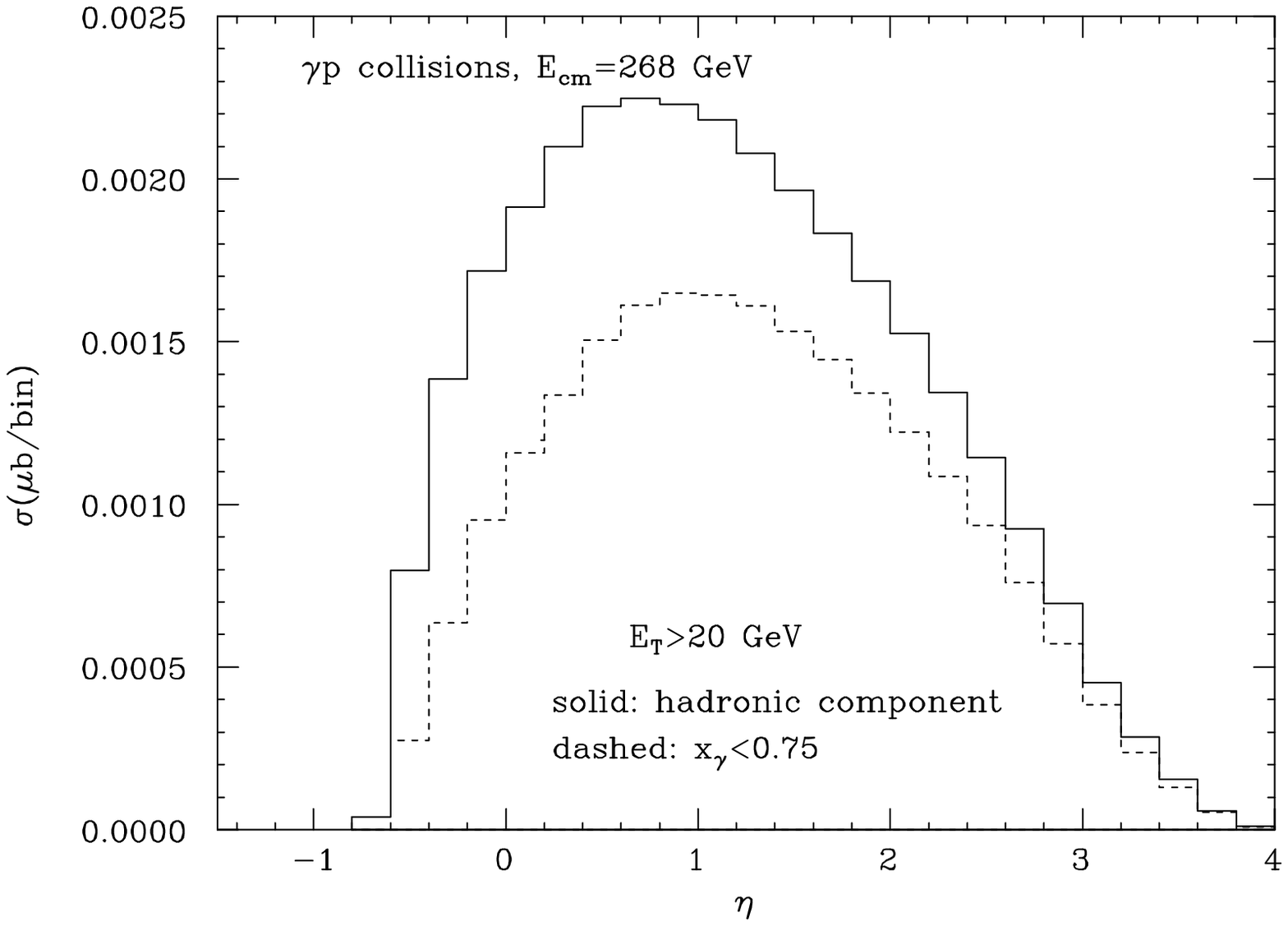,width=0.48\textwidth,clip=} }
\ccaption{}{ \label{fig:eta268}
Pseudorapidity distribution for $x_{\gamma}>0.75$ (left) and 
$x_{\gamma}<0.75$ (right), compared with the pointlike and 
hadronic component respectively. The scales have been set to 
the default value.
}
\end{figure}

We found that the mixing between the hadronic and pointlike components
is sizeable for all the physical observables, even in simple cases
like single-inclusive distributions. As an example, we present in 
fig.~\ref{fig:eta268} the comparison between the hadronic (pointlike) 
component of the $\eta$ distribution, and the corresponding curve 
obtained with a $x_\gamma<0.75$ ($x_\gamma>0.75$) cut. Although in 
this case the shapes are quite similar, the absolute normalization is
rather different.
We have checked that the difference between the two curves
in each plot of fig.~\ref{fig:eta268} is larger than the corresponding
scale uncertainty.

\begin{figure}
\centerline{
   \epsfig{figure=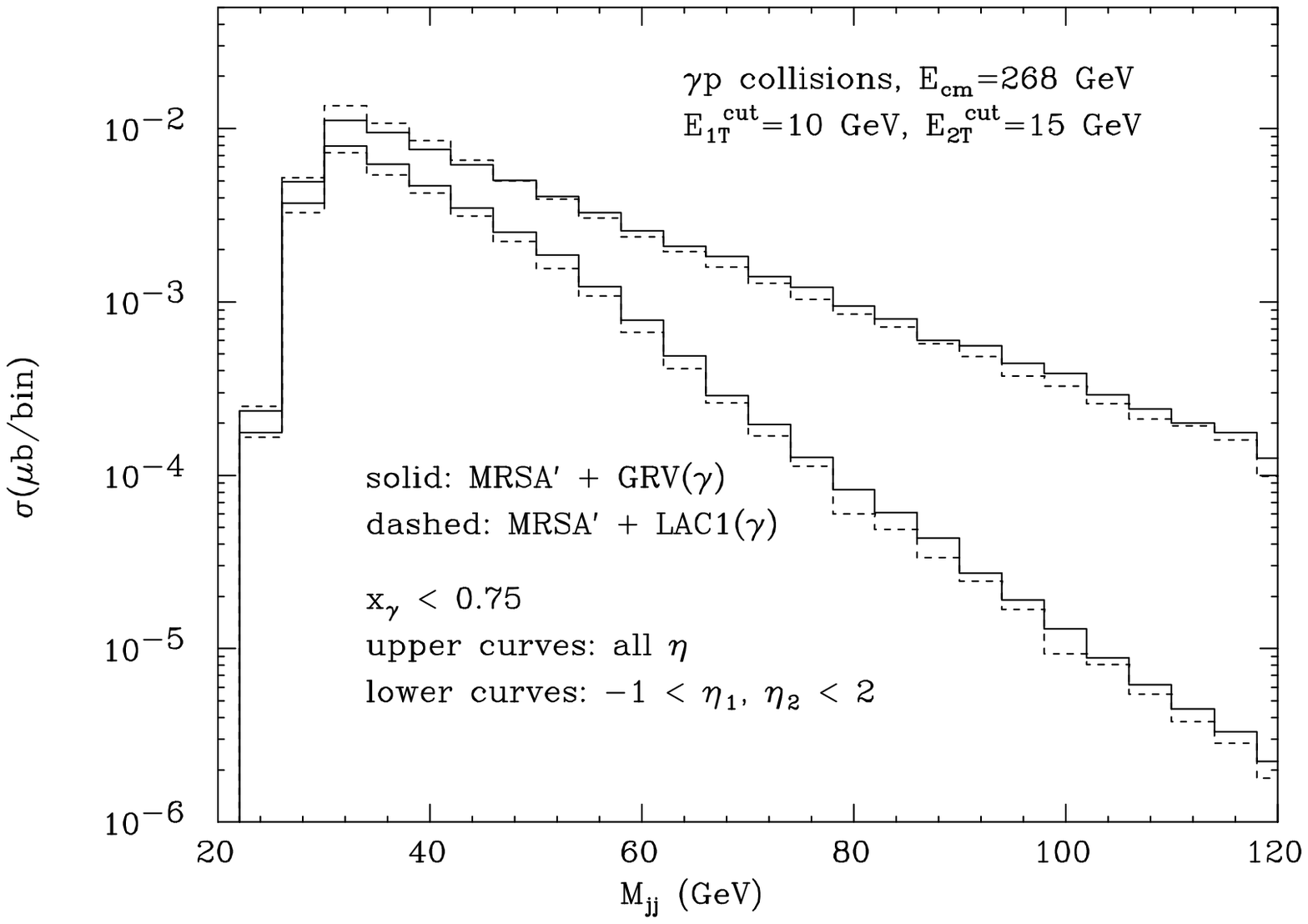,width=0.48\textwidth,clip=}
   \hfill
   \epsfig{figure=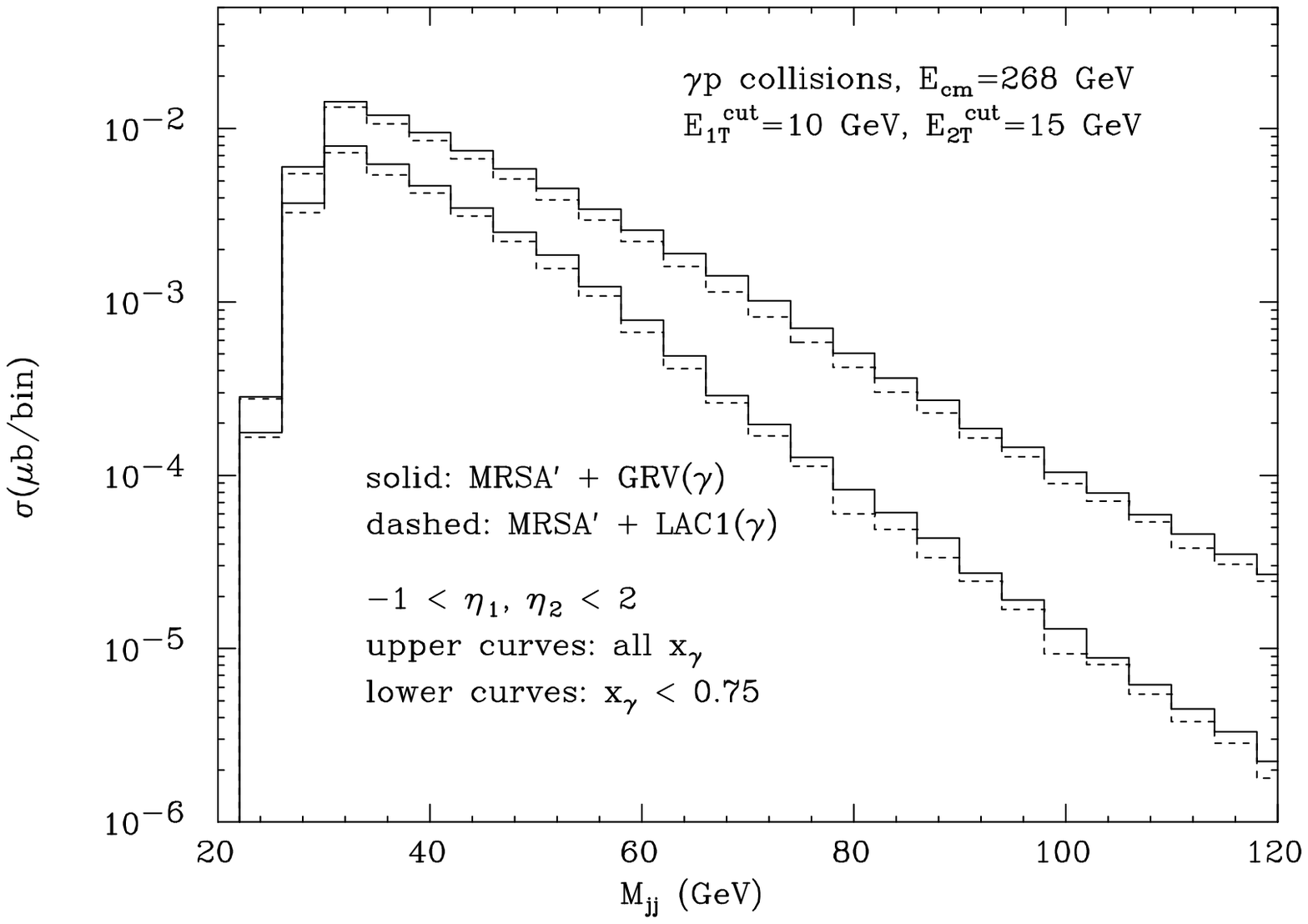,width=0.48\textwidth,clip=} }
\ccaption{}{ \label{fig:mjjstr}
Dependence of the invariant mass distribution upon the parton densities
in the photon.
}
\end{figure}

This discussion suggests that, because of the effect of radiative corrections,
a cut at $x_\gamma=0.75$ is probably not very useful to distinguish between 
partonic subprocesses with or without an incoming photon. Nevertheless, it 
is still conceivable that by cutting in $x_\gamma$, one may select 
observables which are particularly sensitive to the parton densities in 
the photon. In order to further investigate this issue, we have 
computed the invariant mass distribution for $x_\gamma<0.75$ with rather
an extreme choice of the parametrization of parton densities in the photon, 
namely the LAC1~[\ref{LAC}] set, and compared it to our previous 
results obtained with the GRV-HO set. The invariant mass is particularly
suitable to this purpose, since it is directly related to the
Bjorken-$x$ region probed in the collision (for a given center-of-mass
energy, small invariant masses correspond to the small-$x$ region). One 
can see from fig.~\ref{fig:mjjstr} that the difference between the results 
obtained with the GRV and LAC1 sets is comparable to the uncertainty 
band induced by the scale variations we discussed above.
In the central pseudorapidity region the sensitivity to the different
parametrizations is very small. In particular, the right plot of
fig.~\ref{fig:mjjstr} shows that in this $\eta$ region the $x_\gamma$
cut does not enhance the dependence upon the photon parton densities.
Notice that fig.~\ref{fig:mjjstr} is relevant for monochromatic
photon-proton collisions at $E_{cm}=268$~GeV; for smaller center-of-mass
energies or for electron-proton collisions
in the Weizs\"acker-Williams approximation the sensitivity is
even more suppressed, because larger values of the Bjorken-$x$ are probed.
The main difference between the GRV and LAC1 sets is in the behaviour
of the gluon density at small $x$. To effectively probe this region at HERA
smaller values of the invariant mass must be chosen, and QCD predictions 
become less reliable.

\section{Phenomenology of jet production at HERA}
\begin{figure}
\centerline{
   \epsfig{figure=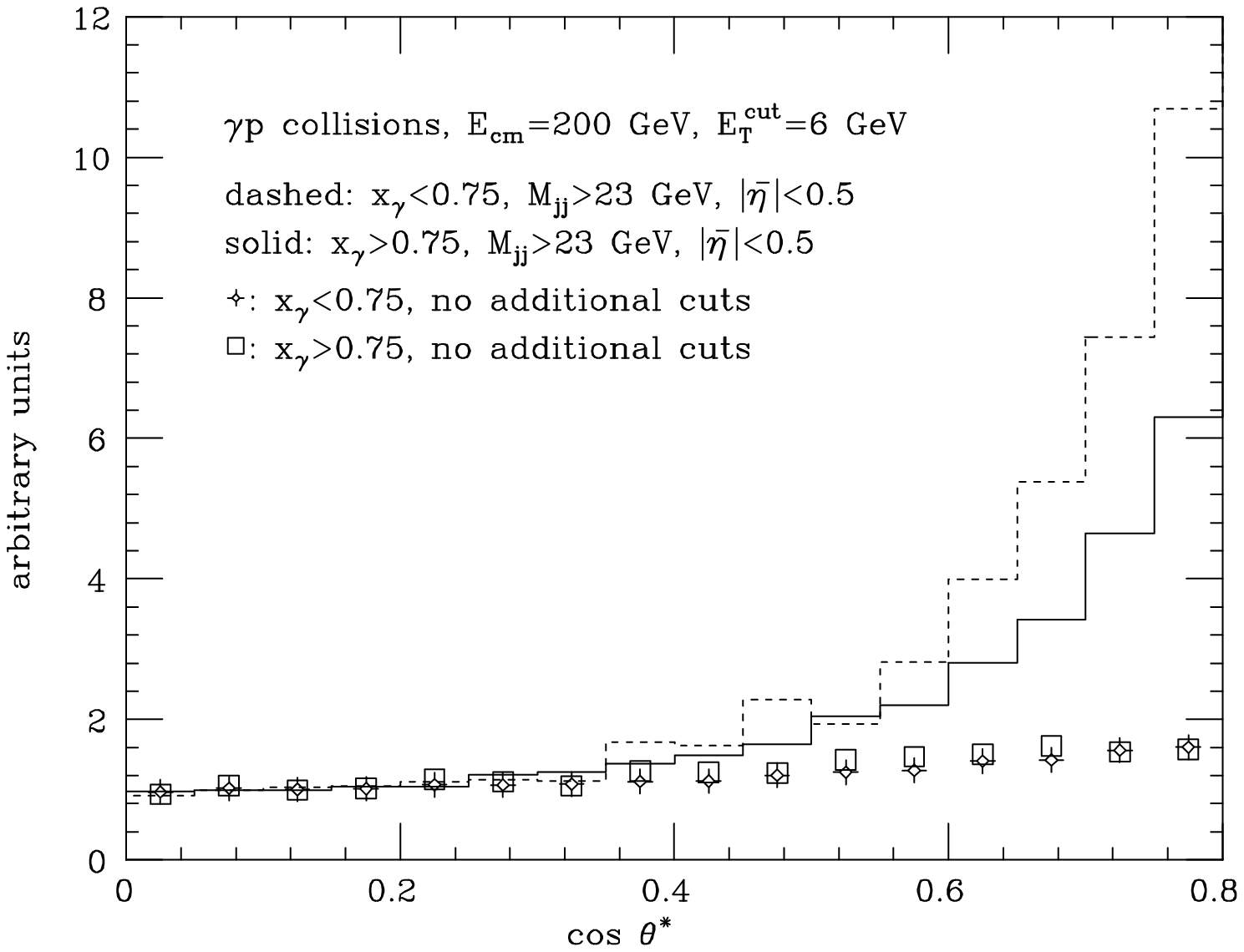,width=0.48\textwidth,clip=}
   \hfill
   \epsfig{figure=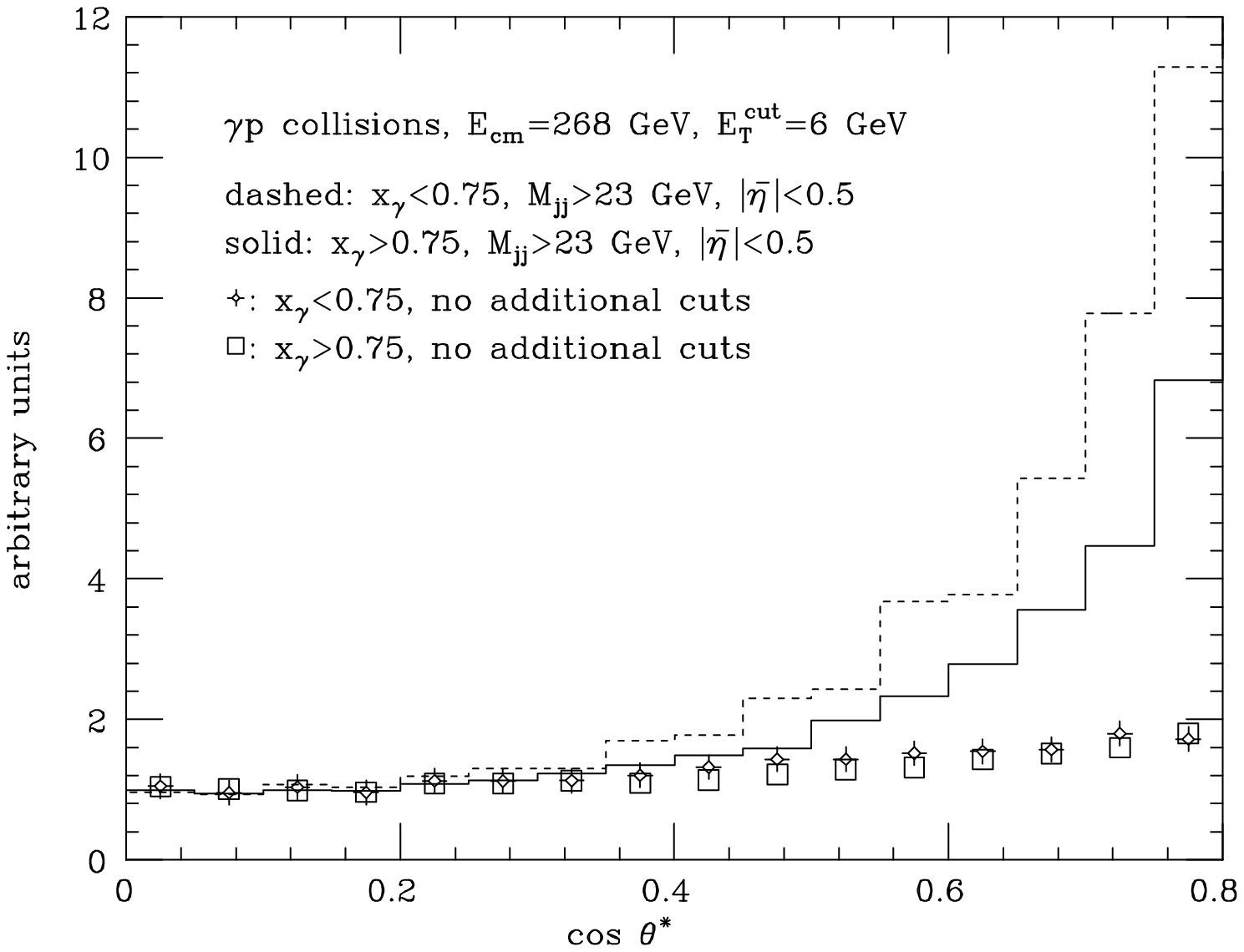,width=0.48\textwidth,clip=} }
\ccaption{}{ \label{fig:ctstar}
Distribution in $\cos\theta^*$, for $\gamma p$ collisions and
various kinematical cuts ($\etonecut=\ettwocut=\etcut$).
}
\end{figure}

In this Section we will discuss the predictions of perturbative QCD
for some quantities of phenomenological interest.
We begin by considering the distribution in the variable $\cos\theta^*$,
defined as
\beq
\cos\theta^*=\tanh\frac{\abs{\eta_1-\eta_2}}{2},
\eeq
where $\eta_{1,2}$ are, as before, the pseudorapidities of the two hardest
jets.
In QCD at leading order the hadronic component
is predicted to behave as $(1-\abs{\cos\theta^*})^{-2}$ when
$\cos\theta^*\to 1$, as opposed to the $(1-\abs{\cos\theta^*})^{-1}$
behaviour of the pointlike component. It is interesting to check whether
the measured distributions for $x_\gamma<0.75$ and $x_\gamma>0.75$ 
display a similar behaviour. In fig.~\ref{fig:ctstar} we present the 
$\cos\theta^*$ distribution in $\gamma p$ collisions for
$\etonecut=\ettwocut=6$~GeV, for two different center-of-mass energies.
We see that for both values of $E_{cm}$ the curves for $x_\gamma>0.75$ 
and $x_\gamma<0.75$, represented in fig.~\ref{fig:ctstar} by squares 
and crosses respectively, have approximately the same shape (the two 
curves have been normalized to have the same integral in the first four 
bins, in order to compare the shapes). This confirms the expectations of 
the previous section, where we have shown that distributions in the 
$x_\gamma<0.75$ (or $x_\gamma>0.75$) region are the result of a significant
mixing between the hadronic and the pointlike components.
It is interesting to notice that a difference in the behaviour
at $\cos\theta^*\sim 1$ of the two distributions obtained for
$x_\gamma<0.75$ and $x_\gamma>0.75$ can be induced by other kinematical 
cuts. To show this, we have computed the $\cos\theta^*$ distribution 
in the two $x_\gamma$ regions, applying the cuts $M_{jj}>23$~GeV 
and $\abs{\bar{\eta}}<0.5$, with $\bar{\eta}=(\eta_1+\eta_2)/2$. These
are the same cuts adopted by the ZEUS collaboration in the analysis of
ref.~[\ref{ZEUSctstar}]. This curves (normalized as before to the same 
integral in the first four bins) are also shown in fig.~\ref{fig:ctstar}; 
we see that the curves for $x_\gamma<0.75$ (dashed) and $x_\gamma>0.75$ 
(solid) display a different behaviour for $\cos\theta^*\to 1$; this 
difference is generated by the introduction of the kinematical cuts. 
This effect can be understood observing that
\beq
M_{jj}^2=2 E_{1\sss T} E_{2\sss T}
\left[ \cosh(\eta_1-\eta_2) - \cos\Delta\phi_{jj}\right].
\eeq
Large values of the invariant mass correspond therefore to large values
of $\cosh(\eta_1-\eta_2)$; on the other hand,
\beq
\cosh(\eta_1-\eta_2)=\frac{1+\cos^2\theta^*}{1-\cos^2\theta^*}.
\label{rr}
\eeq
It is clear from eq.~(\ref{rr}) that a large-$M_{jj}$ cut enhances the region
$\cos\theta^*\sim1$. This shows that the difference in shape between
the solid and dashed curves in fig.~\ref{fig:ctstar} is not directly
related to a different dynamical production mechanism. The curve 
corresponding to $x_\gamma<0.75$ has a steeper rise basically because
the invariant mass distribution relevant for this $x_\gamma$ region
is softer than the invariant mass distribution for $x_\gamma>0.75$.
One may then argue that this is still a signal of different production
mechanisms. But, as we discussed in the previous section, the mixing
between the pointlike and the hadronic components is especially
relevant in the case of $\etonecut=\ettwocut$, and therefore no
particularly significant information on the parton dynamics can be 
extracted from the invariant mass distribution with a $x_\gamma$ cut.

For a given observable, the largest statistics will be collected
by integrating over the energies of the incident photons. For this
reason, we will now present predictions obtained for jet production 
in $ep$ collisions in the Weizs\"acker-Williams approximation,
with $E_{cm}(ep)=300$~GeV, $Q^2=0.01$~GeV$^2$ and 
\mbox{$0.2\leq y\leq 0.8$} (see eqs.~(\ref{wwfun}) and~(\ref{sigep})).
In order to perform a realistic study, we will apply kinematical cuts 
which approximately reproduce the experimental conditions of
the HERA experiments. The pseudorapidity of the observed jets will
be restricted in the range $-1<\eta<2$. In the case of two-jet quantities,
we will require $\etonecut=11$~GeV and $\ettwocut=14$~GeV. By definition,
the two jets are always the hardest of the event.

As discussed in the previous section, with this
choice of cuts on transverse energies
we expect the perturbative expansion to be well-behaved over the whole
phase space. For this reason, we study the scale dependence of our
results by setting all the scales to the same value $\mu$. We present
predictions obtained with two different jet definitions, namely
the cone algorithm, with $R=1$, and the algorithm proposed by Ellis
and Soper in ref.~[\ref{ES}], with $D=1$. In the case
of the cone algorithm we show the full scale variation computed
as before, while for the Ellis-Soper algorithm we only present
the predictions corresponding to our central scale choice.
We recall that a modification of the standard cone algorithm
has been introduced in ref.~[\ref{EKS}] to improve the comparison
with data; this modification however would not change our conclusions,
and we will use \mbox{$R_{sep}=2R$} in the following.

\begin{figure}
\centerline{
   \epsfig{figure=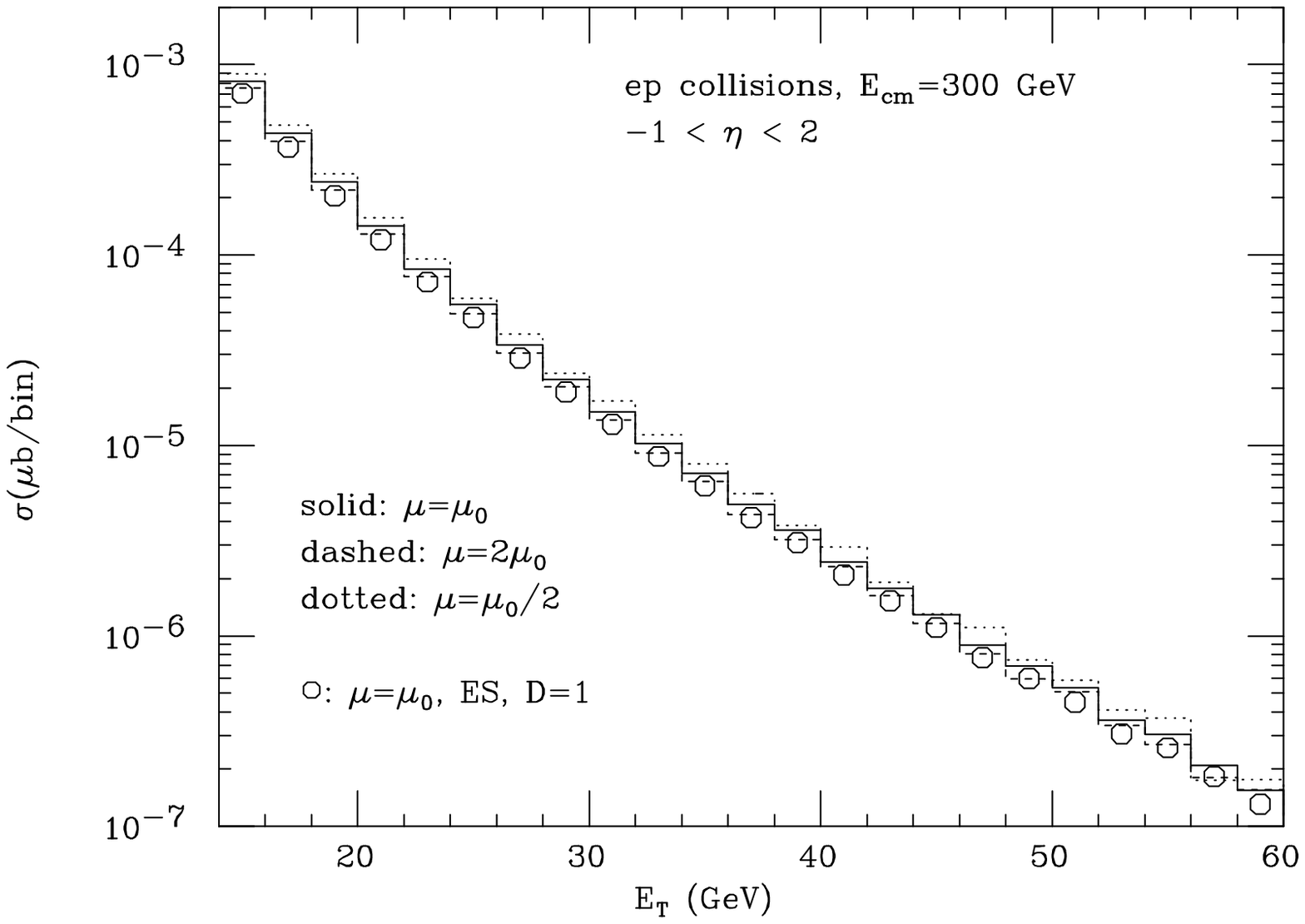,height=0.36\textwidth,
                              width=0.48\textwidth,clip=}
   \hfill
   \epsfig{figure=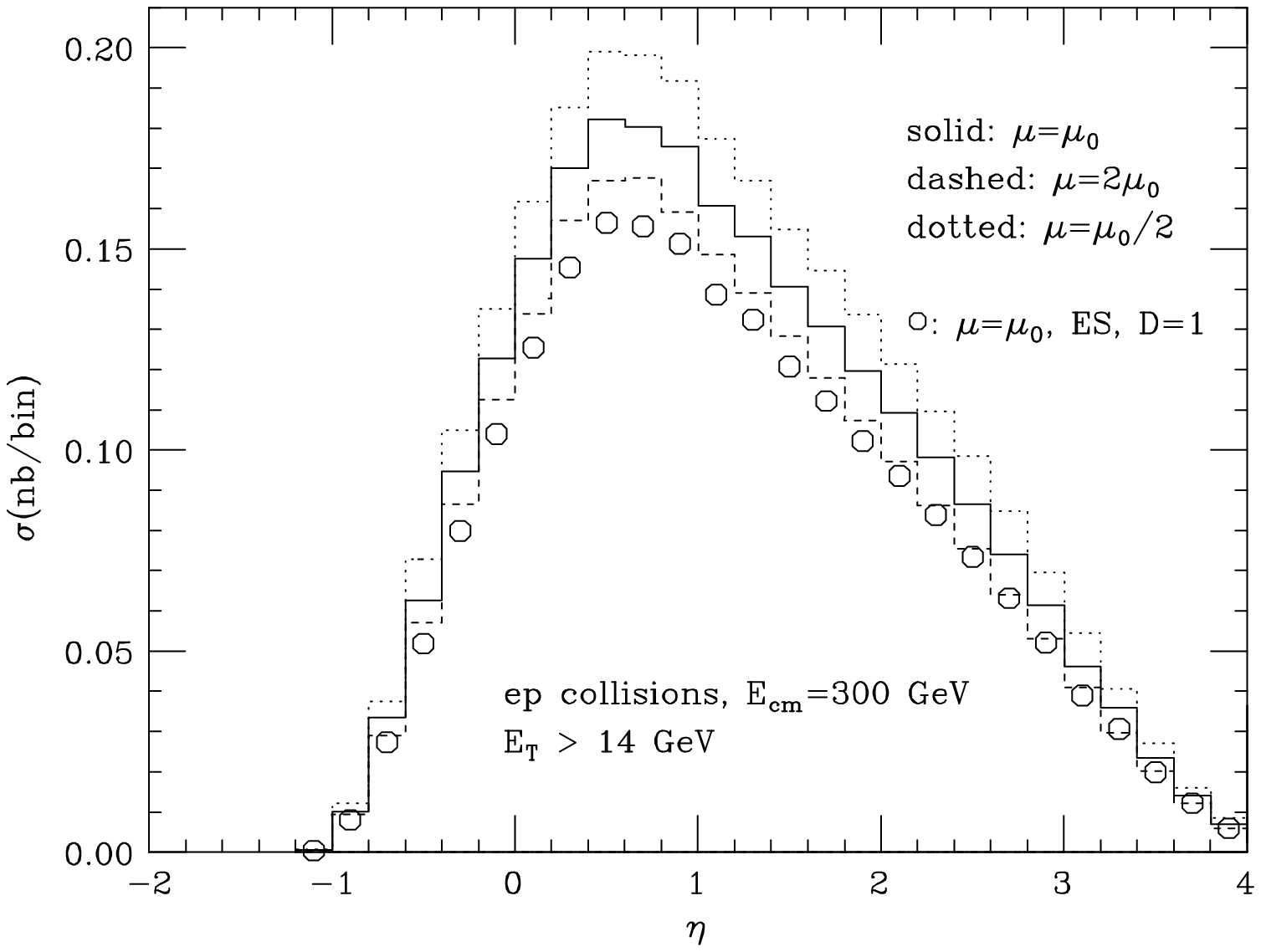,width=0.48\textwidth,clip=} }
\ccaption{}{ \label{fig:pt_eta_h1}
Single-inclusive transverse energy and pseudorapidity. Jets are defined
by the cone algorithm (solid curves; the scale dependence is also shown) 
and by the prescription of ref.~[\ref{ES}] (circles). 
}
\end{figure}

\begin{figure}
\centerline{\epsfig{figure=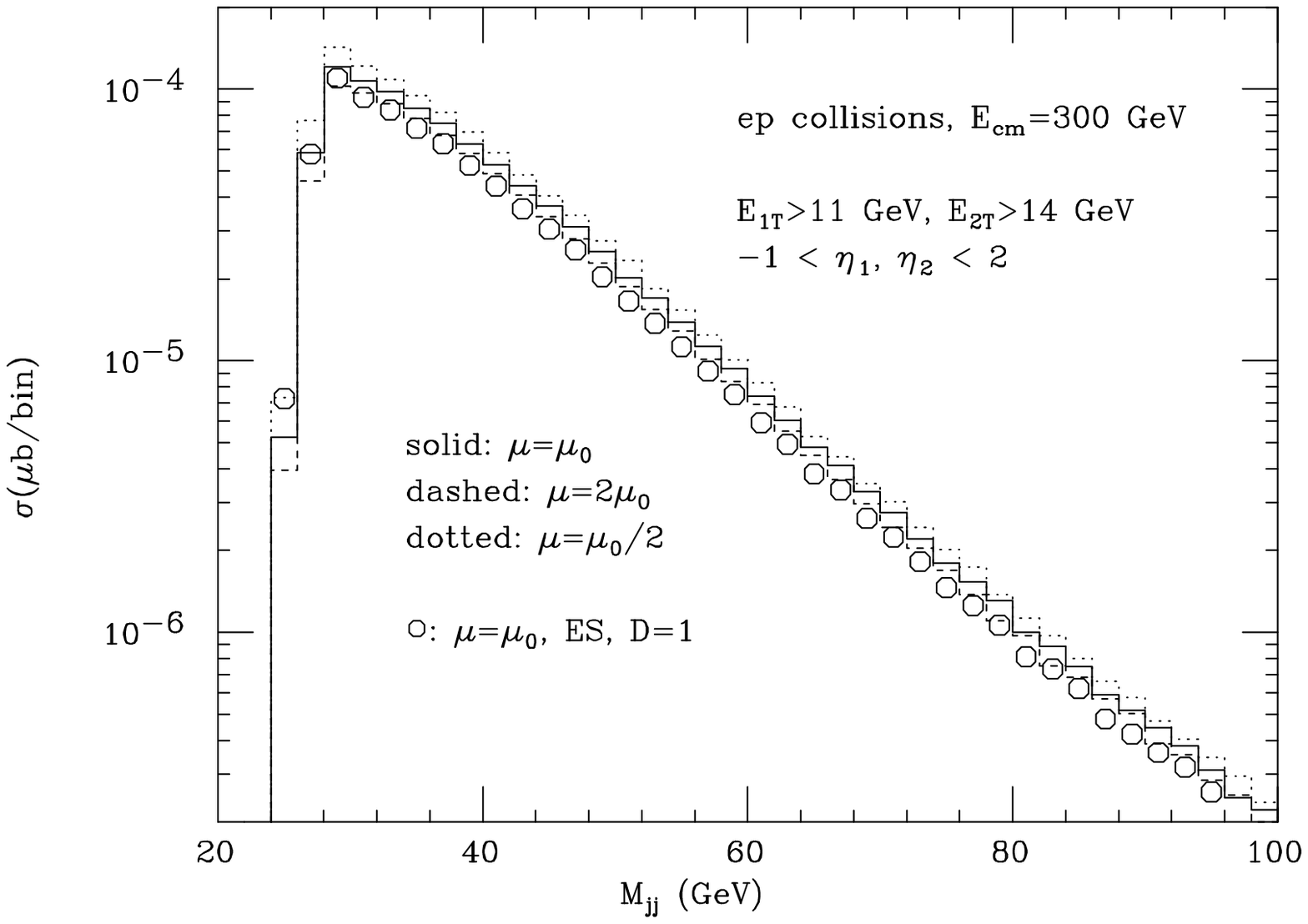,width=0.7\textwidth,clip=}}
\ccaption{}{
\label{fig:mjj_h1}  
Invariant mass of the pair of jets with the largest transverse momenta.
Jets are defined by the cone algorithm (solid curve; the scale dependence 
is also shown) and by the prescription of ref.~[\ref{ES}] (circles). 
}
\end{figure}

\begin{figure}
\centerline{
   \epsfig{figure=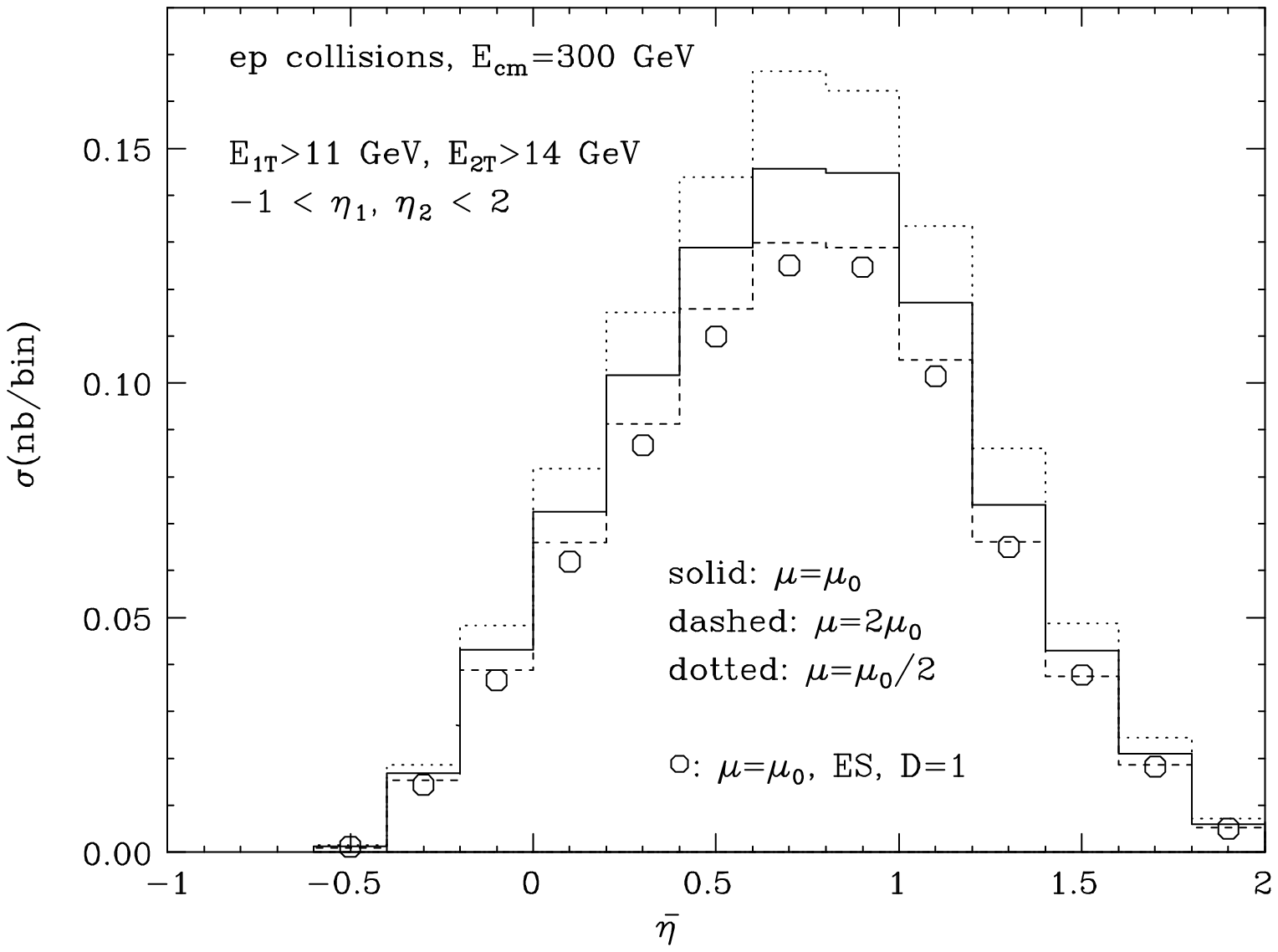,width=0.48\textwidth,clip=}
   \hfill
   \epsfig{figure=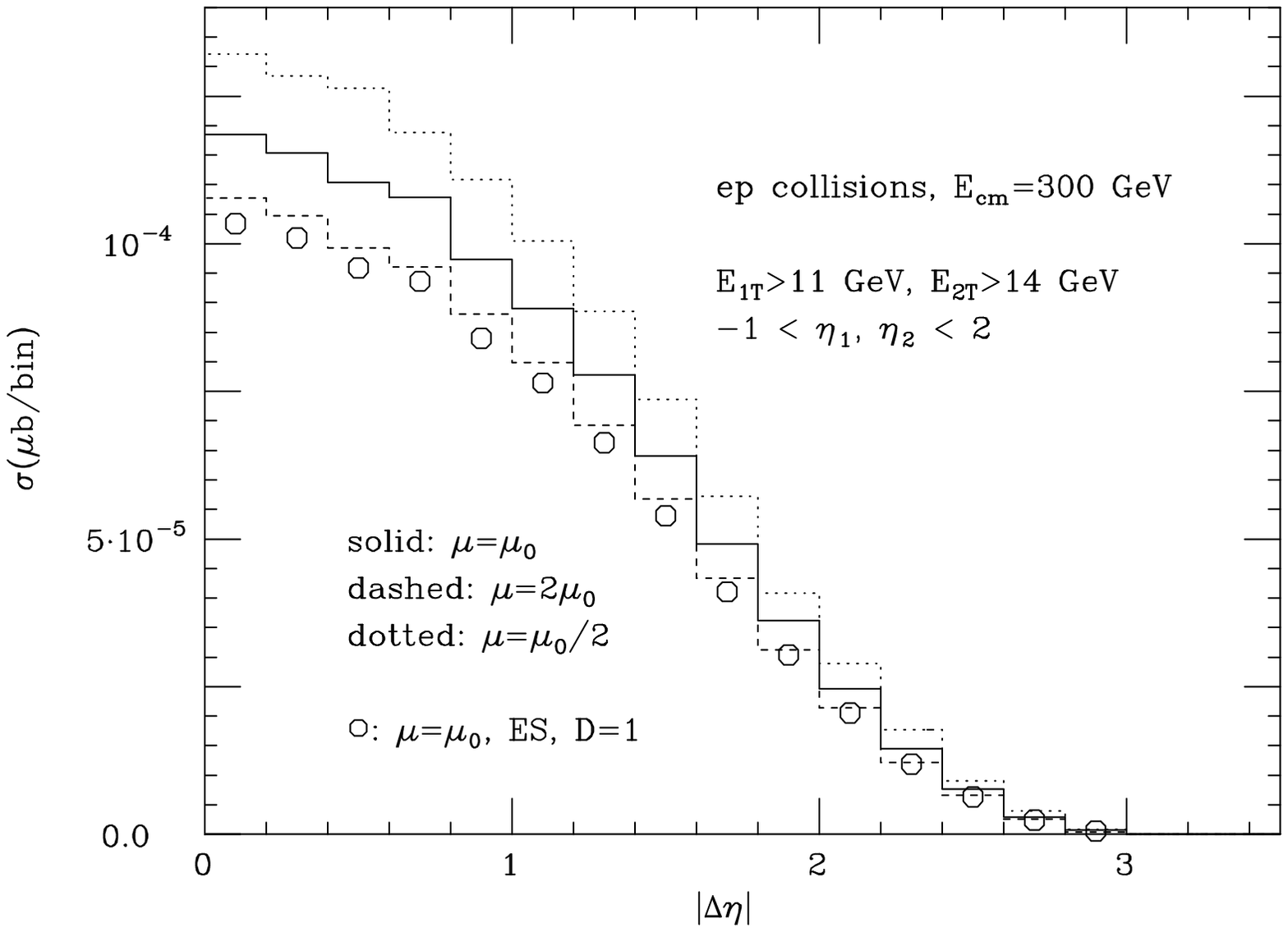,height=0.355\textwidth,
                                width=0.48\textwidth,clip=} }
\ccaption{}{ \label{fig:etabar_deta_h1}
Average pseudorapidity and difference in pseudorapidity 
of the pair of jets with the largest transverse momenta.
Jets are defined by the cone algorithm (solid curves; the scale dependence 
is also shown) and by the prescription of ref.~[\ref{ES}] (circles). 
}
\end{figure}

\begin{figure}
\centerline{
   \epsfig{figure=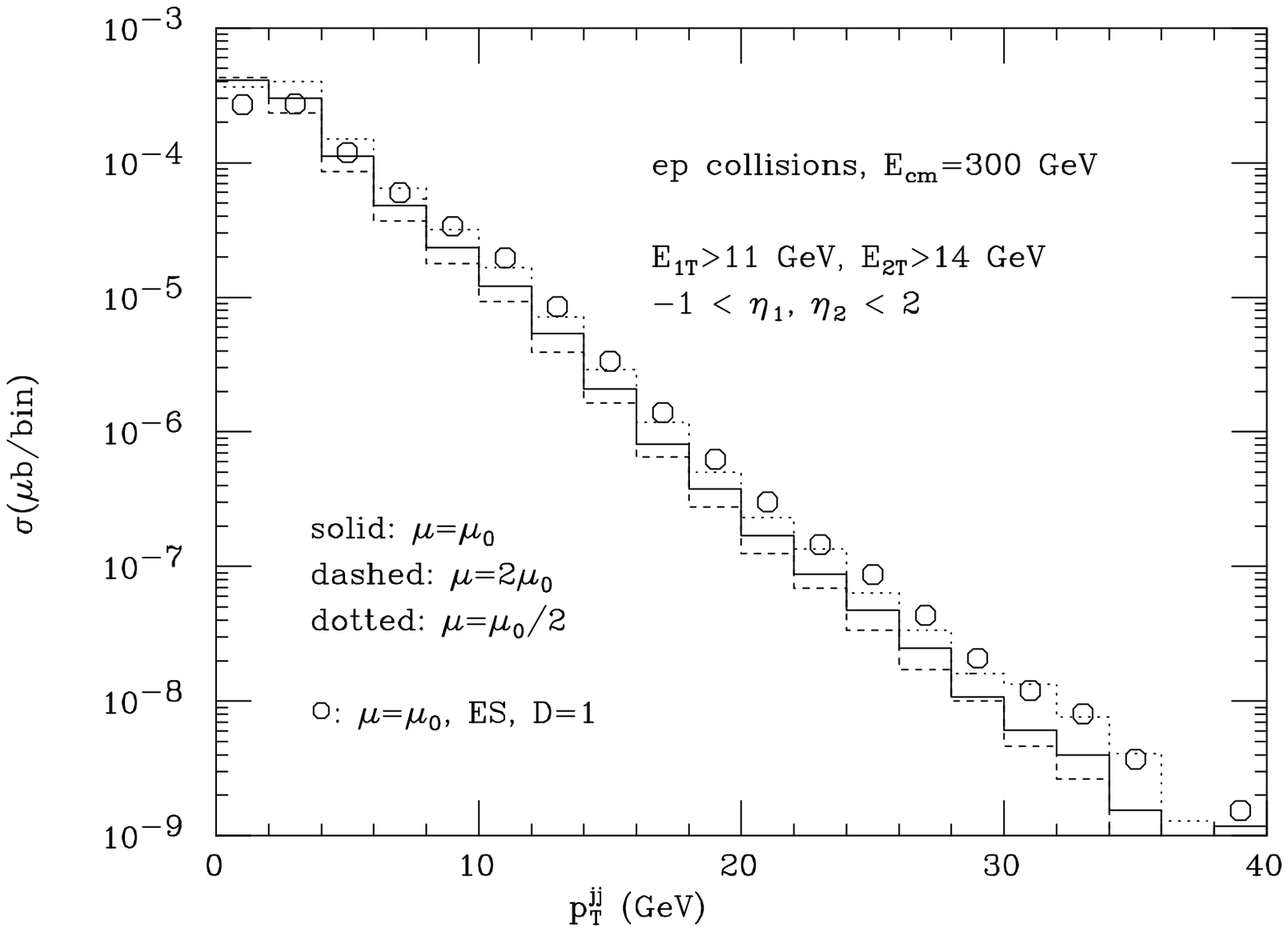,height=0.355\textwidth,
                                width=0.48\textwidth,clip=}
   \hfill
   \epsfig{figure=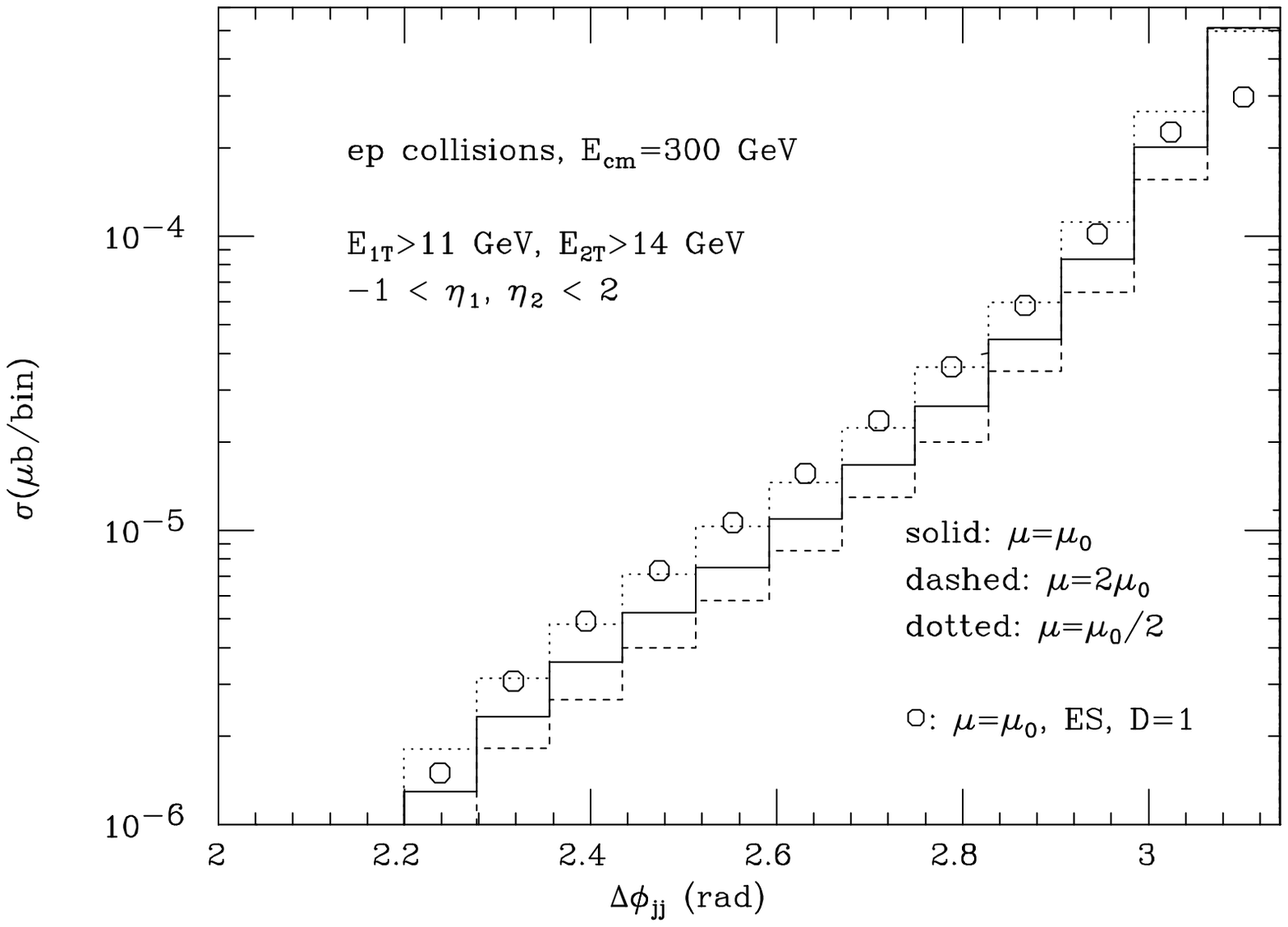,width=0.48\textwidth,clip=} }
\ccaption{}{ \label{fig:ptjj_phi_h1}
Transverse momentum and azimuthal correlation of the pair of jets with 
the largest transverse momenta.
Jets are defined by the cone algorithm (solid curves; the scale dependence 
is also shown) and by the prescription of ref.~[\ref{ES}] (circles). 
}
\end{figure}

In fig.~\ref{fig:pt_eta_h1} we show single-inclusive distributions
in transverse energy and pseudorapidity of the jet (in the latter
case, we require the jet to have transverse energy larger than 14 GeV).
In fig.~\ref{fig:mjj_h1} and fig.~\ref{fig:etabar_deta_h1} we show
two-jet correlations which are non-trivial at leading order:
the invariant mass of the pair, the absolute value of the difference 
in pseudorapidity of the two observed jets, and the average pseudorapidity 
of the pair, $\bar{\eta}=(\eta_1+\eta_2)/2$; notice that the latter 
quantity is equal, at leading order, to the rapidity of the pair.
Finally, fig.~\ref{fig:ptjj_phi_h1} shows correlations trivial
at leading order, the transverse momentum of the pair and
the azimuthal distance. This set of observables gives a fairly complete
description of the production mechanism; it is clearly possible to
consider even more exclusive quantities, by imposing additional
kinematical cuts.

The results presented in figs.~\ref{fig:pt_eta_h1}, \ref{fig:mjj_h1} 
and~\ref{fig:etabar_deta_h1} display a remarkable stability with
respect to scale choice. The difference between the default curve
(solid histogram) and the curves corresponding to $\mu=2\muo$ and 
$\mu=\muo/2$ (dashed and dotted histograms respectively) is
about 10\%. We remark that the non-logarithmic term in the 
Weizs\"acker-Williams function, eq.~(\ref{wwfun}), gives a negative
contribution of the order of 7\% with $Q^2=0.01$~GeV$^2$, and of
5\% with $Q^2=4$~GeV$^2$, and is therefore non-negligible.
Using the jet definition of ref.~[\ref{ES}], represented by
circles, one gets shapes similar to the ones given by the
cone algorithm with $R=0.7$. 

Consistently with what we found in the previous section, the scale
dependence is larger in the case of the plots of fig.~\ref{fig:ptjj_phi_h1},
which are a pure next-to-leading order effect except for the bins
around \mbox{$p_{\sss T}^{jj}=0$} and \mbox{$\Delta\phi_{jj}=\pi$}.
In this case, differences of about 25\% are induced by 
scale variations. The distributions
obtained with the Ellis-Soper algorithm are broader than those
obtained with the cone algorithm and $R=1$, when the jets tend
to be more back-to-back in the transverse plane.

We have also checked that the choice of parton densities does not influence
the QCD predictions very much. For example, using the MRS125 set,
characterized by the fact that $\as(m_Z)$ is fixed to the value 0.125,
the shape of the distributions does not change, and the modification
in the overall normalization is easily traced back to the different
value of $\as$.

\section{Conclusions}
In this paper we have studied the phenomenology of jet 
photoproduction at HERA at next-to-leading order in QCD.
We used the subtraction method, which does not require an
approximate expression for the matrix elements and the
introduction of non-physical infrared cutoffs which are
necessary in other approaches (like the slicing method).
We found that, starting from transverse energy scales of the 
order of $10$~GeV, the theoretical predictions are sufficiently 
stable with respect to scale variation to allow for a significant
comparison with high-statistics data. In some cases,
like two-jet inclusive quantities defined with equal cuts on
the transverse energy of the observed jets, the next-to-leading
order result turns out to be unreliable in some special regions
of the phase space, corresponding typically to configurations
in which the jets are back-to-back in the transverse plane,
or to the threshold of the invariant mass distribution.
In these cases, an all-order resummation would be needed to
get a consistent result. On the other hand, in the remaining
part of the phase space the next-to-leading order result
is well-behaved, and meaningful comparisons with experimental
data can be carried out.

We have considered the problem of the
separation of the pointlike and the hadronic components of the
cross sections. This task could in principle be performed,
if a suitable operational definition could be given for the two
components (the pointlike and hadronic components defined
as in eqs.~(\ref{pointcomp}) and~(\ref{hadrcomp}) are
not physical observables beyond leading order).
We found that by cutting
in the variable $x_\gamma$, defined with the two hardest jets
of a given event, the operationally defined pointlike or 
hadronic components are actually a sizeable mixture between
the next-to-leading order pointlike and hadronic cross sections,
for any reasonable choice of scales.
This implies that this kind of cut is not
very useful to extract information on the underlying parton dynamics.
Also, at the transverse energy scales where the perturbative
expansions is reliable, the sensitivity of the operationally
defined hadronic component to the parton densities in the photon
is rather limited. 

Finally, we have considered a sample set of one-jet
and two-jet inclusive observables which can be measured in order to
have a detailed description of the production mechanism. We have computed
these quantities in the Weizs\"acker-Williams approximation, applying
a set of realistic cuts in order to reproduce the experimental
analyses of the HERA experiments. We found that
the results are remarkably stable with respect to scale variation, 
and therefore that the measurements of these quantities could be used 
as a quality test of perturbative QCD.

\section*{Acknowledgements}
One of us (GR) thanks the Institute of Theoretical Physics of ETH, Zurich,
for the kind hospitality.
We thank C. Grab, Z. Kunszt, H. Niggli, P. Nason and S. Passaggio
for many interesting discussions.
We are indebted to Tancredi Carli, Costas Foudas and Amedeo Staiano
for providing us with useful information about experimental analyses.

\begin{reflist}

\item \label{NLOhadcoll}
 F.~Aversa, M.~Greco, P.~Chiappetta and J.~P.~Guillet, \zp{C46}{90}{253};\\
 \prl{65}{90}{401};\\
 S.~D.~Ellis, Z.~Kunszt and D.~E.~Soper, \prl{64}{90}{2121};\\
 \prl{69}{92}{1496};\\
 W.~T.~Giele, E.~W.~N.~Glover and D.~A.~Kosower, \prl{73}{94}{2019};\\
 S.~D.~Ellis and D.~E.~Soper, \prl{74}{95}{5182}.
\item \label{NLOepemcoll}
 G.~Sterman and S.~Weinberg, \prl{39}{77}{1436};\\
 R.~K.~Ellis, D.~A.~Ross and A.~E.~Terrano, \np{B178}{81}{421};\\
 Z.~Kunszt and P.~Nason, in {\it Z Physics at LEP 1}, eds. G.~Altarelli,
 R.~Kleiss and C.~Verzegnassi, Geneva, 1989;\\
 G.~Kramer and B.~Lampe, {\it Fortschr. Phys.}~{\bf 37}(1989)161;\\
 S.~Bethke, Z.~Kunszt, D.~E.~Soper and W.~J.~Stirling, \np{B370}{92}{310};\\
 W.~T.~Giele and E.~W.~N.~Glower, \pr{D46}{92}{1980};\\
 S.~Catani and M.~H.~Seymour, \pl{B378}{96}{287};\\
 A.~Signer and L.~Dixon, \prl{78}{97}{811}, \hepph{9609460};\\
 preprint SLAC-PUB-7528, \hepph{9706285}.
\item \label{CDFhighet}
 CDF Coll., F.~Abe {\it et al.}, \prl{68}{92}{1104}; \prl{77}{96}{438}.
\item \label{CTEQ4}
 H.~L.~Lai {\it et al.}, \pr{D55}{97}{1280}, \hepph{9606399}.
\item \label{scaling}
 CDF Coll., F.~Abe {\it et al.}, \prl{70}{93}{1376};\\
 see also P.~Melese, for the CDF Coll., preprint FERMILAB-CONF-97-167-E,
 presented at 11$^{th}$ Les Rencontres de Physique de la Vallee d'Aoste, 
 La Thuile, Italy, 2-8 Mar 1997.
\item \label{NLOatHERAnc}
 L.~E.~Gordon and J.~K.~Storrow, \pl{B291}{92}{320};\\
 D. B\"odeker, \zp{C59}{93}{501};\\
 G. Kramer and S.~G.~Salesh, \zp{C61}{94}{277};\\
 D. B\"odeker, G. Kramer and S.~G.~Salesh, \zp{C63}{94}{471};\\
 M.~Klasen and G.~Kramer, \zp{C72}{96}{107}, \hepph{9511405}.
\item \label{KKfull}
 M.~Klasen and G.~Kramer, preprint DESY 96-246, \hepph{9611450}.
\item \label{HO}
 B.~W.~Harris and J.~F.~Owens, preprint FSU-HEP-970411, \hepph{9704324}. 
\item \label{mcjet}
 S.~Frixione, preprint ETH-TH/97-14, \hepph{9706545}.
\item \label{Butterworth}
 J.~M.~Butterworth, for the H1 and ZEUS Coll., preprint UCL/HEP 97-04, 
 presented at the Ringberg Workshop, Tegernsee, 25-30 May 1997,
 \hepex{9707001}.
\item \label{Aurenche}
 P.~Aurenche, preprint ENSLAPP-A-652-97, talk given at Photon 97, 
 Egmond aan Zee, Netherlands, 10-15 May 1997, \hepph{9706386}.
\item\label{FKS}
 S.~Frixione, Z.~Kunszt and A.~Signer, \np{B467}{96}{399}, \hepph{9512328}.
\item\label{WW}
 C.F.~Weizs\"acker, \zp{88}{34}{612}; E.J.~Williams, \pr{45}{34}{729}.
\item\label{WW2}
 V.M.~Budnev {\it et al.}, \pr{C15}{74}{181};\\
 H.A.~Olsen, \pr{D19}{79}{100}.
\item\label{FMNRWW}
 S.~Frixione, M.~Mangano, P.~Nason and G.~Ridolfi, \pl{B319}{93}{339}.
\item \label{conealg}  
 F.~Aversa {\it et al.}, Proceedings of the Summer Study on 
 High Energy Physics, Research Directions for the Decade, 
 Snowmass, CO, Jun 25 - Jul 13, 1990.
\item \label{MRSAprime}  
 A.~D.~Martin, R.~G.~Roberts and W.~J.~Stirling, \pl{B354}{95}{155},
  \hepph{9502336}.
\item \label{GRVPDFpho}  
 M.~Gl\"uck, E.~Reya and A.~Vogt, \pr{D46}{92}{1973}.
\item\label{FMNRPF}
 S.~Frixione, M.L.~Mangano, P.~Nason and G.~Ridolfi, \np{B412}{94}{225}.
\item\label{LAC}
 H.~Abramowicz, K.~Charchula and A.~Levy, \pl{B269}{91}{458}.
\item\label{ZEUSctstar}
 M.~Derrick {\it et al.}, ZEUS Coll., \pl{B384}{96}{401}, \hepex{9605009}.
\item\label{ES}
 S.~D.~Ellis and D.~E.~Soper, \pr{D48}{93}{3160}.
\item\label{EKS}
 S.~D.~Ellis, Z.~Kunszt and D.~E.~Soper, \prl{69}{92}{3615}.

\end{reflist}

\end{document}